# Removing Propagation Redundant Constraints in Redundant Modeling

C.W. CHOI and J.H.M. LEE
The Chinese University of Hong Kong
and
P. J. STUCKEY
NICTA and University of Melbourne

A widely adopted approach to solving constraint satisfaction problems combines systematic tree search with various degrees of constraint propagation for pruning the search space. One common technique to improve the execution efficiency is to add redundant constraints, which are constraints logically implied by others in the problem model. However, some redundant constraints are *propagation redundant* and hence do not contribute additional propagation information to the constraint solver. Redundant constraints arise naturally in the process of redundant modeling where two models of the same problem are connected and combined through channeling constraints. In this paper, we give general theorems for proving propagation redundancy of one constraint with respect to channeling constraints and constraints in the other model. We illustrate, on problems from CSPlib (`http://www.csplib.org/`), how detecting and removing propagation redundant constraints in redundant modeling can significantly speed up constraint solving.

## 1. INTRODUCTION

Finite domain constraint programming combines backtracking tree search with constraint propagation to solve *constraint satisfaction problems* (CSPs) [Mackworth 1977]. *Constraint propagation* removes infeasible values from the domains of variables to reduce the search space. This propagation-based constraint solving framework is realized in modern constraint programming systems such as ECL$^i$PS$^e$ [Cheadle et al. 2003], ILOG Solver [1999], and SICStus Prolog [2003], which have been successfully applied to many real-life industrial applications.

There is usually more than one way of modeling a problem as a CSP. By modeling a problem as a CSP, we mean the process of determining the variables, the associated domains of the variables, and the expressions of the constraints. Finding a good model of a CSP is a challenging task. A modeler must specify a set of constraints that capture the definitions of the problem, but this is not enough. The model should also have *strong* propagation: that is, it should be able to quickly reduce the domains of the variables of the problem. Moreover, the implementation of *propagators* to perform constraint propagation should be efficient. Last but not least, the model should have a small search space.[1]

A common technique to increase propagation strength is to add *redundant constraints*, which are logically implied by the constraints of the model. Adding redundant constraints can be beneficial since the constraint solver may extract more information from these redundant constraints. However, some logically redundant constraints are *propagation redundant*, and hence do not contribute additional propagation information to the constraint solver. Generally, we only want to add redundant constraints that are propagation non-redundant to reduce the search space.

An important source of logically redundant constraints is in *redundant modeling* [Cheng et al. 1999]. A problem can be modeled differently from two viewpoints using two different sets of variables. By connecting the two different models with *channeling constraints*, which relate valuations in the two different models, stronger propagation behavior can be achieved in the combined model. However, the additional variables and constraints impose extra computation overhead. Given each model is complete and only admits the solutions of the problem then each model is logically redundant with respect to the other model plus the channeling constraints. In many cases, some of the constraints are also propagation redundant with respect to the other constraints in the combined model. By reasoning about propagation redundancy, we can improve redundant modeling by just keeping the constraints which give beneficial *new* propagation.

In this paper, we introduce the notion of *restrictive and unrestrictive channel functions* to characterize channeling constraints. We study the propagation behavior of constraints based on the notion of *propagation rules*, which capture each possible propagation by a constraint. This allows us to systematically determine if a propagator is redundant with respect to the propagators of a set of constraints through simple implication tests. We give general theorems for proving propagation redundancy of constraints involved in redundant models. We focus on propagators that perform (the combination of) two popular propagation techniques, namely *domain propagation* [Van Hentenryck et al. 1998] and *set bounds propagation* [Gervet 1997]. We illustrate, on problems from CSPLib (http://www.csplib.org/), how detecting and removing propagation redundant constraints can significantly speed up solving behavior. This paper is a revised and extended version of our earlier work [Choi et al. 2003a; 2003b].

The remainder of the paper is organized as follows. In Section 2 we introduce propagation-based constraint solving and *propagation rules*, a way of enumerating the different propagation behaviors of a propagator. In Section 3 we define a broad form of channeling constraints that are covered by our approach. In Section 4 we give theorems that allow us to show which constraints in a redundant model are not causing extra propagation and can be removed. In Section 5 we discuss some important issues in studying about propagation redundancy. In Section 6 we give experimental results showing the benefits of detecting and removing propagation redundant constraints. In Section 7 we discuss related work. In Section 8 we summarize our contributions and shed light on future directions of research.

---

[1] For example, the search space of a problem model using integer variables is usually smaller than that using Boolean variables.

## 2. BACKGROUND

In this paper we consider integer and set constraint solving with constraint propagation and tree search. Boolean constraint solving is considered as a special case of integer constraint solving. Our notations, although different from the conventional CSP literatures, allow us to express the theoretical framework in a simpler manner.

### 2.1 Variables and Domains

We consider a typed set of variables $\mathcal{V} = \mathcal{V}_I \cup \mathcal{V}_S$ made up of *integer* variables $\mathcal{V}_I$, for which we use lower case letters such as $x$ and $y$, and *sets of integers* variables $\mathcal{V}_S$, for which we use upper case letters such as $S$ and $T$. We use $v$ to denote variables of either kind.

Each variable is associated with a finite set of possible values, defined by the domain of the CSP. A *domain* $D$ is a complete mapping from a fixed (countable) set of variables $\mathcal{V}$ to finite sets of integers (for the integer variables in $\mathcal{V}_I$) and to finite sets of finite sets of integers (for the set variables in $\mathcal{V}_S$). A *false domain* $D$ is a domain with $D(v) = \emptyset$ for some $v$. A *singleton* domain $D$ is such that $|D(v)| = 1$ for all $v \in \mathcal{V}$. The *intersection* of two domains $D_1$ and $D_2$, denoted $D_1 \sqcap D_2$, is defined by the domain $D_3(v) = D_1(v) \cap D_2(v)$ for all $v$. A domain $D_1$ is *stronger* than a domain $D_2$, written $D_1 \sqsubseteq D_2$, if $D_1(v) \subseteq D_2(v)$ for all variables $v$. A domain $D_1$ is equal to a domain $D_2$, denoted $D_1 = D_2$, if $D_1(v) = D_2(v)$ for all variables $v$. We also use *range* notation whenever possible: $[\,l\,..\,u\,]$ denotes the set $\{d \mid l \leq d \leq u\}$ when $l$ and $u$ are integers, while $[\,L\,..\,U\,]$ denotes the set of sets of integers $\{A \mid L \subseteq A \subseteq U\}$ when $L$ and $U$ are sets of integers. We shall be interested in the notion of an *initial domain*, which we denote $D_{init}$. The initial domain gives the initial values possible for each variable. In effect an initial domain allows us to restrict attention to domains $D$ such that $D \sqsubseteq D_{init}$.

### 2.2 Valuations, Infima and Suprema

A *valuation* $\theta$ is a mapping of integer variables ($x_i \in \mathcal{V}_I$) to integer values and set variables ($S_i \in \mathcal{V}_S$) to sets of integer values, written $\{x_1 \mapsto d_1, \ldots, x_n \mapsto d_n, S_1 \mapsto A_1, \ldots, S_m \mapsto A_m\}$ where $d_i \in D(x_i)$ and $A_j \in D(S_j)$. Let *vars* be the function that returns the set of variables appearing in an expression, constraint or valuation. Given an expression $e$, $\theta(e)$ is obtained by replacing each $v \in vars(e)$ by $\theta(v)$ and calculating the value of the resulting variable free expression. In an abuse of notation, we define a valuation $\theta$ to be an element of a domain $D$, written $\theta \in D$, if $\theta(v_i) \in D(v_i)$ for all $v_i \in vars(\theta)$.

Define the *infimum* and *supremum* of an expression $e$ with respect to a domain $D$ as $\inf_D e = \inf\{\theta(e) | \theta \in D\}$ and $\sup_D e = \sup\{\theta(e) | \theta \in D\}$. The ordering $\preceq$

used by inf and sup depends on the type of the expression. If $e$ has integer type then $d_1 \preceq d_2$ iff $d_1 \leq d_2$, while if $e$ has set of integer type then $d_1 \preceq d_2$ iff $d_1 \subseteq d_2$. Note that these values may not exist for arbitrary domains and set of integer type expressions. Later we shall restrict ourselves to domains and expression where infimum and supremum always do exist.

## 2.3 Constraints

A constraint places restriction on the allowable values for a set of variables and is usually written in well understood mathematical syntax. More formally, a *constraint* $c$ is a relation expressed using available function and relation symbols in a specific constraint language. For the purpose of this paper, we assume the usual integer interpretation of arithmetic constraints, set operators such as $\in$ and $\subseteq$, and logical operators such as $\neg, \wedge, \vee, \Rightarrow$, and $\Leftrightarrow$. A CSP consists of a set of constraints read as conjunction. We use *true* to denote the empty (always satisfiable) conjunction of constraints. We define $solns(c) = \{\theta \mid vars(\theta) = vars(c) \wedge \models_\theta c\}$, that is the set of $\theta$ that make the constraint $c$ hold true. We call $solns(c)$ the *solutions* of $c$. In some cases, constraints can also be defined directly by giving the set (or table) $solns(c)$. We sometimes treat an integer constraint $c$ as an expression with value 1 if true and 0 if false. We can understand a domain $D$ as a constraint in the obvious way, $D \leftrightarrow \bigwedge_{v \in \mathcal{V}} \bigvee_{d \in D(v)} v = d$. A constraint $c$ is *logically redundant* with respect to a constraint $c'$ if $\models c' \rightarrow c$, that is $c$ holds whenever $c'$ holds.

## 2.4 Propagators and Propagation Solvers

In the context of propagation-based constraint solving, a constraint specifies a *propagator*, which gives the basic units of propagation. A *propagator* $f$ is a monotonically decreasing function from domains to domains, i.e. $D_1 \sqsubseteq D_2$ implies that $f(D_1) \sqsubseteq f(D_2)$, and $f(D) \sqsubseteq D$. A propagator $f$ is *correct* for constraint $c$ iff for all domains $D$

$$\{\theta \mid \theta \in D\} \cap solns(c) = \{\theta \mid \theta \in f(D)\} \cap solns(c)$$

This is a weak restriction since for example, the identity propagator is correct for all constraints $c$. We will usually assume that a propagator $f$ for constraint $c$ is *checking*, that is, if $D$ is a singleton domain, then $f(D) = D$ iff $\exists \theta \in D \cap solns(c)$. A checking propagator correctly determines the satisfiability of the constraint $c$ for singleton domains.

A *propagation solver* for a set of propagators $F$ and current domain $D$, $solv(F, D)$, repeatedly applies all the propagators in $F$ starting from domain $D$ until there is no further change in resulting domain. $solv(F, D)$ is the largest domain $D' \sqsubseteq D$ which is a fixpoint (i.e. $f(D') = D'$) for all $f \in F$. In other words, $solv(F, D)$ returns a new domain defined by

$$iter(F, D) = \sqcap_{f \in F} f(D) \quad \text{and} \quad solv(F, D) = \text{gfp}(\lambda d.iter(F, d))(D)$$

where gfp denotes the greatest fixpoint w.r.t $\sqsubseteq$ lifted to functions.

## 2.5 Domain and Set Bounds Propagators

Propagators are often (but not always) linked to implementing some notion of *local consistency*. The most well-studies consistency notion is *arc consistency* [Mackworth 1977] which ensures that for each constraint, every value in the domain of the first variable, has a supporting value in the domain of the second variable which satisfied the constraint. Arc consistency can be naturally extended to constraints of more than two variables. This extension has been called *generalized arc consistency* [Mohr and Masini 1998], as well as *domain consistency* [Van Hentenryck et al. 1998] (which is the terminology we will use), and *hyper-arc consistency* [Marriott and Stuckey 1998].

A domain $D$ is *domain consistent* for a constraint $c$ if $D$ is the least domain containing all solutions $\theta \in D$ of $c$, i.e, there does not exist $D' \sqsubset D$ such that $\theta \in D \wedge \theta \in solns(c) \rightarrow \theta \in D'$. A set of propagators $F$ implement *domain consistency* for a constraint $c$, if $solv(F, D)$ is always domain consistent for $c$. Define the *domain propagator* for a constraint $c$ as

$$dom(c)(D)(v) = \begin{cases} \{\theta(v) \mid \theta \in D \wedge \theta \in solns(c)\} & \text{where } v \in vars(c) \\ D(v) & \text{otherwise} \end{cases}$$

EXAMPLE 2.1. Consider the constraint $c \equiv x_1 = 3x_2 + 5x_3$. Suppose domain $D(x_1) = \{2, 3, 4, 5, 6, 7\}$, $D(x_2) = \{0, 1, 2\}$, and $D(x_3) = \{-1, 0, 1, 2\}$. The solutions $\theta \in D$ of $c$ are $\theta_1 = \{x_1 \mapsto 3, x_2 \mapsto 1, x_3 \mapsto 0\}$, $\theta_2 = \{x_1 \mapsto 5, x_2 \mapsto 0, x_3 \mapsto 1\}$, and $\theta_3 = \{x_1 \mapsto 6, x_2 \mapsto 2, x_3 \mapsto 0\}$. Hence, $dom(c)(D) = D'$ where $D'(x_1) = \{3, 5, 6\}$, $D'(x_2) = \{0, 1, 2\}$, and $D'(x_3) = \{0, 1\}$. $D'$ is domain consistent with respect to $c$. □

Set bounds propagation [Gervet 1997] is typically used where a domain maps a set variable to a lower bound set of integers and an upper bound set of integers. We shall enforce this by restricting our attention to domains where the $D(S)$ is a range, that is $D(S) = \{A \mid \inf_D(S) \subseteq A \subseteq \sup_D(S)\}$. This is managed by only using set bounds propagators, which maintain this property. The set bounds propagator returns the smallest set range which includes the result returned by the domain propagator. Define the *set bounds propagator* for a constraint $c$ where $vars(c) \subseteq \mathcal{V}_S$ as

$$sb(c)(D)(v) = \begin{cases} [\cap dom(c)(D)(v) \,..\, \cup dom(c)(D)(v)] & \text{where } v \in vars(c) \\ D(v) & \text{otherwise} \end{cases}$$

EXAMPLE 2.2. Consider the constraint $c \equiv S_1 \subseteq S2$, suppose the domain $D(S_1) = [\,\{1\}\,..\,\{1, 2, 3, 4\}\,]$, $D(S_2) = [\,\emptyset\,..\,\{1, 2, 3\}\,]$. Then, $D' = sb(c)(D)$ where $D'(S_1) = D'(S_2) = [\,\{1\}\,..\,\{1, 2, 3\}\,]$. □

A constraint can involve both integer and set variables. We define the *domain and set bounds propagator* $dsb(c)$ for a constraint $c$ as follows:

$$dsb(c)(D)(v) = \begin{cases} sb(c)(D)(v) & \text{where } v \in vars(c) \cap \mathcal{V}_S \\ dom(c)(D)(v) & \text{otherwise} \end{cases}$$

Note that as defined $dsb(c) = dom(c)$ when $vars(c) \subseteq \mathcal{V}_I$. From now on we shall restrict attention to $dsb$ propagators.

EXAMPLE 2.3. Consider the domain propagator $dom(|S| = x)$, applied to a domain $D$ where $D(x) = \{2\}$, and $D(S) = [\emptyset .. \{1,5,8\}]$. Then the solutions $\theta \in D$ are $\{x \mapsto 2, S \mapsto \{1,5\}\}$, $\{x \mapsto 2, S \mapsto \{1,8\}\}$, $\{x \mapsto 2, S \mapsto \{5,8\}\}$. So $D' = dom(|S| = x)(D)$ gives $D'(S) = \{\{1,5\}, \{1,8\}, \{5,8\}\}$. The domain and set bounds propagator $dsb(|S| = x)$ instead determines $D = dsb(|S| = x)(D)$ since $\cap\{\{1,5\}, \{1,8\}, \{5,8\}\} = \emptyset$ and $\cup\{\{1,5\}, \{1,8\}, \{5,8\}\} = \{1,5,8\}$. □

## 2.6 Atomic Constraints and Propagation Rules

An atomic constraint represents the basic changes in domain that occur during propagation: the elimination of a value from an integer domain, or the addition of a value to a lower bound, or removal of a value from an upper bound. An *atomic constraint* is one of $x_i = d$, $x_i \neq d$, $d \in S_i$ or $d \notin S_i$ where $x_i \in \mathcal{V}_I$, $d$ is an integer, and $S_i \in \mathcal{V}_S$. [2]

A *propagation rule* is of the form $C \rightarrowtail c$ where $C$ is a conjunction of *atomic constraints*, $c$ is an atomic constraint, and $\not\models C \rightarrow c$. For notational convenience we shall write extended rules $C \rightarrowtail C'$ where $C'$ is a conjunction of atomic constraints as a shorthand for the set of rules $\{C \rightarrowtail c \mid c \in C'\}$.

A propagation rule $C \rightarrowtail c$ defines a propagator (for which we use the same notation) in the obvious way.

$$(C \rightarrowtail c)(D)(v) = \begin{cases} \{\theta(v) \mid \theta \in D \wedge \theta \in solns(c)\} & \text{if } vars(c) = \{v\} \text{ and } \models D \rightarrow C \\ D(v) & \text{otherwise} \end{cases}$$

We can characterize an arbitrary propagator $f$ in terms of the propagation rules that it implements. A propagator $f$ *implements* a propagation rule $C \rightarrowtail c$ if for each $D \sqsubseteq D_{init}$ whenever $\models D \rightarrow C$, then $\models f(D) \rightarrow c$. Let $rules(f)$ be the set of rules implemented by $f$. This definition of $f$ is often unreasonably large. In order to reason more effectively about propagation rules for a given propagator, we need to have a minimal representation.[3] Let $prop(f) \subseteq rules(f)$ be a set of propagation rules such that $solv(prop(f), D) = solv(rules(f), D)$ for all $D \sqsubseteq D_{init}$, and there does not exist a $\overline{prop}(f) \subset prop(f)$ for which $solv(\overline{prop}(f), D) = solv(rules(f), D)$ for all $D \sqsubseteq D_{init}$. That is, all propagation caused by $f$ is also caused by the *minimal set* of propagation rules $prop(f)$. Notice that $prop(f)$ is not unique.

EXAMPLE 2.4. For Boolean constraint $c \equiv z_{12} + z_{13} + z_{14} = 1$ where $D_{init}(z_{12}) = D_{init}(z_{13}) = D_{init}(z_{14}) = [0..1]$ the propagation rules $prop(dsb(c))$ are

$$\begin{array}{ll} z_{12} = 1 \rightarrowtail z_{13} = 0, z_{14} = 0 & z_{12} = 0, z_{13} = 0 \rightarrowtail z_{14} = 1 \\ z_{13} = 1 \rightarrowtail z_{12} = 0, z_{14} = 0 & z_{12} = 0, z_{14} = 0 \rightarrowtail z_{13} = 1 \\ z_{14} = 1 \rightarrowtail z_{12} = 0, z_{13} = 0 & z_{13} = 0, z_{14} = 0 \rightarrowtail z_{12} = 1 \end{array}$$

Note that propagation rules $prop(f)$ for propagators $f$ for Boolean constraints need only ever involve equations since $\models D_{init} \rightarrow (z = b \leftrightarrow z \neq 1 - b)$ for $b \in \{0, 1\}$. □

A *key lemma* for domain and set bounds propagators $dsb(c')$, is that the propagation rules implemented are exactly those $C \rightarrowtail c$ where $c'$ implies $C \rightarrow c$.

---

[2] Atomic constraints of the form $x_i = d$ are not strictly necessary for propagation rules. They are equivalent to removing all other values from the domain.
[3] Both Brand [2003] and Abdennadher and Rigotti [2002] give effective methods for creating minimal representations of any constraints in terms of propagation rules.

LEMMA 2.5. *Given a constraint $c'$, $dsb(c')$ implements $C \rightarrowtail c$ iff $\models (D_{init} \wedge c') \rightarrow (C \rightarrow c)$.*

PROOF. ($\Rightarrow$) Suppose $dsb(c')$ implements $C \rightarrowtail c$. Then for each $D \sqsubseteq D_{init}$ we have that $\models D \rightarrow C$ means that $\models dsb(c')(D) \rightarrow c$.

Suppose to the contrary $\not\models (D_{init} \wedge c') \rightarrow (C \rightarrow c)$. Thus there exists solution $\theta \in D_{init}$ that satisfies $c' \wedge C \wedge \neg(c)$. We build a domain $D_\theta$ from a valuation $\theta$ as follows. $D_\theta(v) = \{\theta(v)\}$ for all $v \in vars(\theta)$, while $D_\theta(v) = D_{init}(v)$ otherwise. Now $\models D_\theta \rightarrow C$ and $dsb(c')(D_\theta) = D_\theta$ since $\theta$ is a solution of $c'$. But then $\not\models dsb(c')(D_\theta) \rightarrow c$. Contradiction.

($\Leftarrow$) Suppose $\models (D_{init} \wedge c') \rightarrow (C \rightarrow c)$. Then for any domain $D \sqsubseteq D_{init}$ where $\models D \rightarrow C$, we have that $\models (D \wedge c') \rightarrow c$ since $\models D \rightarrow D_{init}$. Hence, every solution $\theta \in D$ of $c'$ is a solution of $c$.

Suppose to the contrary $dsb(c')$ does not implement $C \rightarrowtail c$. Then there exists $D \sqsubseteq D_{init}$, such that $\models D \rightarrow C$ and $D' = dsb(c')(D)$ but $\not\models D' \rightarrow c$. We now prove a contradiction for each form of $c$:

— If $c \equiv d \in S$, then $\not\models D' \rightarrow c$ means $d \notin \inf_{D'}(S)$. Hence by the definition of $dsb(c')$, there is a solution $\theta \in D$ of $c'$ where $d \notin \theta(S)$. Contradiction since $\theta$ is not a solution of $c$.
— If $c \equiv d \notin S$, then $\not\models D' \rightarrow c$ means $d \in \sup_{D'}(S)$. Hence by the definition of $dsb(c')$ there is a solution $\theta \in D$ of $c'$ where $d \in \theta(S)$. Contradiction since $\theta$ is not a solution of $c$.
— If $c \equiv x \neq d$, then $\not\models D' \rightarrow c$ means $d \in D'(x)$. Hence by the definition of $dsb(c')$, there is a solution $\theta \in D$ of $c_S$ where $\theta(x) = d$. Contradiction since $\theta$ is not a solution of $c$.
— If $c \equiv x = d$, then $\not\models D' \rightarrow c$ means $d' \in D'(x)$ where $d' \neq d$. Hence by the definition of $dsb(c')$, there is a solution $\theta \in D$ of $c'$ where $\theta(x) = d'$. Contradiction since $\theta$ is not a solution of $c$.

□

## 3. CHANNELING CONSTRAINTS

Redundant modeling [Cheng et al. 1999] models a problem from more than one viewpoint. By joining two models using channeling constraints, we can get the advantage of both sources of propagation.

Assume we have one model of the problem $M_X$ using variables $X$, and another model $M_Y$ using disjoint variables $Y$. Channeling constraints can be used to join these two models together by relating $X$ and $Y$. There is no real agreement, as yet, as to precisely what channeling constraints are. For the purposes of our theorems we define a channeling constraint as follows.

Let $A_X$ be the atomic constraints for $D_{init}$ on variables $X$, and $A_Y$ be the atomic constraints for $D_{init}$ on variables $Y$. A *channel function* $\Diamond$ is a bijection from atomic constraints $A_X$ to $A_Y$.

A *channeling constraint* (or simply *channel*) $C_\Diamond$ is the constraint

$$\bigwedge_{c \in A_X} (c \Leftrightarrow \Diamond(c))$$

The *channel propagator* $F_\diamond$ is the set of propagation rules inferred from the channel function $\diamond$.

$$F_\diamond = \bigcup_{c \in A_X} \{c \rightarrowtail \diamond(c), \diamond(c) \rightarrowtail c\}$$

Note for channel function $\diamond$, by definition $\diamond^{-1}$ is also a channel function, and $C_\diamond$ and $C_{\diamond^{-1}}$, as well as $F_\diamond$ and $F_{\diamond^{-1}}$, are identical.

We now illustrate how common channels fit into this framework.

### 3.1 Permutation Channels

A common form of redundant modeling is when we consider two viewpoints to a *permutation problem* [Geelen 1992]. In a permutation problem, the objective is to find a bipartite matching between two sets of objects $A = \{a_1, \ldots, a_n\}$ and $B = \{b_1, \ldots, b_n\}$ satisfying all other problem specific constraints. Generally, we can model a permutation problem from two different viewpoints. In the first viewpoint, we assign objects from $B$ to $A$. We use the set of variables $X = \{x_1, \ldots, x_n\}$ to denote objects in $A$, and the domain $D(x_i) = \{1, \ldots, n\}$, for all $1 \leq i \leq n$ to denote objects in $B$. The second viewpoint swaps the role between $A$ and $B$, i.e. assign objects from $A$ to $B$. We use the set of variables $Y = \{y_1, \ldots, y_n\}$ to denote objects in $B$, and the domain $D(y_j) = \{1, \ldots, n\}$, for all $1 \leq j \leq n$ to denote objects in $A$.

The *permutation channel function* $\bowtie$ is defined as $\bowtie(x_i = j) = (y_j = i)$ and $\bowtie(x_i \neq j) = (y_j \neq i)$ for all $1 \leq i, j \leq n$. The *permutation channel* $C_\bowtie$ is equivalent to the conjunction of constraints

$$\bigwedge_{i=1}^{n} \bigwedge_{j=1}^{n} (x_i = j \Leftrightarrow y_j = i)$$

EXAMPLE 3.1. **Langford's Problem** The problem "prob024" in CSPLib is an example of permutation problem. The problem is to find an $(m \times n)$-digit sequence which includes the digits 1 to $n$, with each digit occurring $m$ times. There is one digit between any consecutive pair of the digit 1, two digits between any consecutive pair of the digit 2, ..., and $n$ digits between any consecutive pair of the digit $n$.

Smith [2000] suggests two ways to model the Langford's problem. We use the $(3 \times 9)$ instance to illustrate the two models. In the first model, $M_X$, we use 27 variables $X = \{x_1, \ldots, x_{27}\}$, which we can think of as $1_1, 1_2, 1_3, 2_1, \ldots, 9_2, 9_3$. Here, $1_1$ represents the first digit 1 in the sequence, $1_2$ represents the second digit 1, and so on. The initial domain of these variables, $D_{init}(x_i) = \{1, \ldots, 27\}$ for $1 \leq i \leq 27$, represents the positions of the digit $x_i$ in the sequence. We enlist the constraints of Smith's model as follows:

—(LX1) disequality constraints: $\forall 1 \leq i < j \leq 27.\ x_i \neq x_j$
—(LX2.1) separation constraints: $\forall 1 \leq i \leq 9.\ x_{3i-1} = x_{3i-2} + (i+1)$
—(LX2.2) separation constraints: $\forall 1 \leq i \leq 9.\ x_{3i} = x_{3i-1} + (i+1)$

In the second model, $M_Y$, we again use 27 variables $Y = \{y_1, \ldots, y_{27}\}$ to represent each position in the sequence. The initial domain of these variables, $D_{init}(y_i) = \{1, \ldots, 27\}$ for $1 \leq i \leq 27$, corresponds to the digits $1_1, 1_2, 1_3, 2_1, \ldots, 9_2, 9_3$ in position $y_i$ of the sequence. The constraints are:

—(LY1) disequality constraints: $\forall 1 \leq i < j \leq 27.\ y_i \neq y_j$

—(LY2.1) separation constraints: $\forall 1 \leq i \leq 9. \forall 1 \leq j \leq 27 - 2(i+1).\ y_j = 3i - 2 \Leftrightarrow y_{j+(i+1)} = 3i - 1$

—(LY2.2) separation constraints: $\forall 1 \leq i \leq 9. \forall 1 \leq j \leq 27 - 2(i+1).\ y_j = 3i - 2 \Leftrightarrow y_{j+2(i+1)} = 3i$

—(LY3) separation constraints: $\forall 1 \leq i \leq 9. \forall (28 - 2(i+1)) \leq j \leq 27.\ y_j \neq 3i - 2$

The permutation channel for the two models is simply $x_i = j \Leftrightarrow y_j = i$ for all $1 \leq i, j \leq 27$. □

EXAMPLE 3.2. **All Interval Series Problem** The problem "prob007" in CSPLib is from musical composition. The problem is to find a permutation of $n$ numbers from 1 to $n$, such that the differences between adjacent numbers form a permutation from 1 to $n - 1$. We give two ways to model the problem. The first model slightly modifies the model suggested by Puget and Régin [2001], and the the second model slightly modifies the model suggested by Choi and Lee [2002].

The first model, $M_X$, consists of $n$ variables, $X = \{x_1, \ldots, x_n\}$. Each $x_i$ denotes the number in position $i$, and $D_{init}(x_i) = [1..n]$ for $1 \leq i \leq n$. We introduce auxiliary variables, $\{u_1, \ldots, u_{n-1}\}$ that denote the difference between adjacent numbers, where $D_{init}(u_i) = [1..n-1]$ for $1 \leq i \leq n - 1$. The constraints are:

—(IX1.1) disequality constraints: $\forall 1 \leq i < j \leq n.\ x_i \neq x_j$

—(IX1.2) disequality constraints: $\forall 1 \leq i < j \leq n - 1.\ u_i \neq u_j$

—(IX2) interval constraints: $\forall 1 \leq i \leq n - 1.\ u_i = |x_i - x_{i+1}|$

The second model, $M_Y$, also consists of $n$ variables, $Y = \{y_1, \ldots, y_n\}$. Each $y_i$ denotes the position for the number $i$, and $D_{init}(y_i) = [1..n]$ for $1 \leq i \leq n$. The auxiliary variables $\{v_1, \ldots, v_{n-1}\}$ denote the position where the difference value of 1 to $n - 1$ belongs, and $D_{init}(v_i) = [1..n-1]$ for $1 \leq i \leq n - 1$. The constraints are:

—(IY1.1) disequality constraints: $\forall 1 \leq i < j \leq n.\ y_i \neq y_j$

—(IY1.2) disequality constraints: $\forall 1 \leq i < j \leq n - 1.\ v_i \neq v_j$

—(IY2.1) interval constraints: $\forall 1 \leq i < j \leq n.\ (y_i - y_j = 1) \Rightarrow (v_{j-i} = y_j)$

—(IY2.2) interval constraints: $\forall 1 \leq i < j \leq n.\ (y_j - y_i = 1) \Rightarrow (v_{j-i} = y_i)$

The (IY2.1) and (IY2.2) constraints enforce that if $y_i$ and $y_j$ are adjacent, the position for their difference must be the smaller of them. In the second model, observe the fact that only the numbers 1 and $n$ can give us the difference of $n - 1$. Therefore, we can add the following redundant constraints:

$$(\text{IY3}): (|y_1 - y_n| = 1) \wedge (v_{n-1} = \min(y_1, y_n)),$$

which requires $y_1$ and $y_n$ to be adjacent.

The permutation channels for this problem are more interesting because we have two distinct kinds of variables in each model, each of which is related by a permutation channel. The channels are $x_i = j \Leftrightarrow y_j = i$ for all $1 \leq i, j \leq n$ and $u_i = j \Leftrightarrow v_j = i$ for all $1 \leq i, j \leq n - 1$. □

### 3.2 Boolean Channels

Another common form of redundant modeling is when we give both an integer and Boolean models. Suppose we have an integer model using the integer variables $X = \{x_1, \ldots, x_n\}$ and the domain $D_{init}(x_i) = [1..k]$. We can have a corresponding Boolean model using the Boolean variables $Z = \{z_{ij} \mid 1 \leq i \leq n, 1 \leq j \leq k\}$. Each variable $z_{ij}$ encodes the proposition that $x_i = j$.

The *Boolean channel function* $\triangle$ is defined as $\triangle(x_i = j) = (z_{ij} = 1)$ and $\triangle(x_i \neq j) = (z_{ij} = 0)$ for all $1 \leq i \leq n, 1 \leq j \leq k$. Note that the atomic constraints $z_{ij} \neq 1$ and $z_{ij} \neq 0$ are not needed for Boolean variables since they are equivalent (respectively) to $z_{ij} = 0$ and $z_{ij} = 1$. The *Boolean channel* $C_\triangle$ is equivalent to the conjunction of constraints

$$\bigwedge_{i=1}^{n} \bigwedge_{j=1}^{k} (x_i = j \Leftrightarrow z_{ij} = 1)$$

EXAMPLE 3.3. **n-Queens Problem** The problem is to place $n$ queens on an $n \times n$ chess board so that no two queens can attack each other. There are two common ways to model this problem, i.e., an integer model and a Boolean model.

The integer model, $M_X$, consists of $n$ variables, $X = \{x_1, \ldots, x_n\}$. Each $x_i$ denotes the column position of the queen on row $i$, and $D_{init}(x_i) = \{1, \ldots, n\}$, for $1 \leq i \leq n$. The constraints are:

—(QX1) column constraints: $\forall 1 \leq i < j \leq n.\ x_i \neq x_j$
—(QX2.1) diagonal constraints: $\forall 1 \leq i < j \leq n.\ x_i - i \neq x_j - j$
—(QX2.2) diagonal constraints: $\forall 1 \leq i < j \leq n.\ x_i + i \neq x_j + j$

The Boolean model, $M_Z$, consists of $n \times n$ Boolean variables, $Z = \{z_{11}, \ldots, z_{1n}, \ldots, z_{n1}, \ldots, z_{nn}\}$. Each Boolean variable $z_{ij}$ denotes whether we have a queen at row $i$ column $j$ or not. The constraints are:

—(QZ1) row constraints: $\forall 1 \leq i \leq n.\ \sum_{j=1}^{n} z_{ij} = 1$
—(QZ2) column constraints: $\forall 1 \leq j \leq n.\ \sum_{i=1}^{n} z_{ij} = 1$
—(QZ3.1) main diagonal constraint: $\sum_{i=1}^{n} z_{ii} \leq 1$
—(QZ3.2) main diagonal constraint: $\sum_{i=1}^{n} z_{i(n-i+1)} \leq 1$
—(QZ4.1) diagonal constraints: $\forall 1 \leq k \leq n-1.\ \sum_{j=1}^{n-k} z_{j(j+k)} \leq 1$
—(QZ4.2) diagonal constraints: $\forall 1 \leq k \leq n-1.\ \sum_{j=1}^{n-k} z_{(j+k)j} \leq 1$
—(QZ4.3) diagonal constraints: $\forall 1 \leq k \leq n-1.\ \sum_{j=1}^{n-k} z_{j(n-j-k+1)} \leq 1$
—(QZ4.4) diagonal constraints: $\forall 1 \leq k \leq n-1. \sum_{j=1}^{n-k} z_{(j+k)(n-j+1)} \leq 1$

We combine the two models using the Boolean channel $x_i = j \Leftrightarrow z_{ij} = 1$ for all $1 \leq i \leq n, 1 \leq j \leq k$.  □

### 3.3 Set Channels

Another common form of redundant modeling is where one model deals with integer variables, and the other with variables over finite sets of integers, and the relation $x_i = j$ holds iff $i \in S_j$. This generalizes the permutation problem to where two or more integer variables can take the same value. Suppose the integer variables are $X = \{x_1, \ldots, x_n\}$, where $D_{init}(x_i) = [\,1 .. k\,]$ for all $1 \leq i \leq n$, and the set variables are $S = \{S_1, \ldots, S_k\}$ where $D_{init}(S_j) = [\,\emptyset .. \{1, \ldots, n\}\,]$ for all $1 \leq j \leq k$.

The *set channel function* $\{\}$ is defined as $\{\}(x_i = j) = (i \in S_j)$ and $\{\}(x_i \neq j) = (i \notin S_j)$ for all $1 \leq i \leq n, 1 \leq j \leq k$. The *set channel* $C_{\{\}}$ is equivalent to

$$\bigwedge_{i=1}^{n} \bigwedge_{j=1}^{k} (x_i = j \Leftrightarrow i \in S_j)$$

EXAMPLE 3.4. **Social Golfers Problem** The problem "prob010" in CSPLib is to arrange $n = g \times s$ players into $g$ groups of $s$ players each week, playing for $w$ weeks, so that no two players play in the same group twice. Smith [2001] suggests two ways to model this problem.

In the first model we use variables $X = \{x_{lk} | 1 \leq l \leq n, 1 \leq k \leq w\}$ to denote the group which player $l$ plays on week $k$, and $D_{init}(x_{lk}) = [\,1 .. g\,]$ for all $1 \leq l \leq n, 1 \leq k \leq w$.

The constraints of the problem are expressed as:

—(GX1) each group has $s$ players: $\forall 1 \leq i \leq g. \forall 1 \leq k \leq w.\ \Sigma_{l=1}^{n}(x_{lk} = i) = s$
—(GX2) two players only play in the same group in one week:

$$\forall 1 \leq k_1 \neq k_2 \leq w. \forall 1 \leq l_1 \neq l_2 \leq n.\ \neg(x_{l_1 k_1} = x_{l_2 k_1} \wedge x_{l_1 k_2} = x_{l_2 k_2})$$

The second model uses set variables $S = \{S_{ik} | 1 \leq i \leq g, 1 \leq k \leq w\}$ to denote the set of players play in group $i$ on week $k$. and $D_{init}(S_{ik}) = [\,\emptyset .. \{1, \ldots, n\}\,]$ for all $1 \leq i \leq g, 1 \leq k \leq w$. The constraints are expressed as:

—(GS1) no groups in the same week have a player in common:

$$\forall 1 \leq k \leq w. \forall 1 \leq i_1 \neq i_2 \leq g.\ S_{i_1 k} \cap S_{i_2 k} = \emptyset$$

—(GS2) each group has $s$ players: $\forall 1 \leq i \leq g. \forall 1 \leq k \leq w.\ |S_{ik}| = s$
—(GS3) no different groups have more than one player in common:

$$\forall 1 \leq i_1 \neq i_2 \leq g. \forall 1 \leq k_1 \neq k_2 \leq w.\ |S_{i_1 k_1} \cap S_{i_2 k_2}| \leq 1$$

We can use the set channels to combine the two models, $x_{lk} = i \Leftrightarrow l \in S_{ik}$ for all $1 \leq l \leq n, 1 \leq k \leq w, 1 \leq i \leq g$. □

EXAMPLE 3.5. **Balanced Academic Curriculum Problem** The problem, listed as "prob030" in CSPLib, is to design an academic curriculum aiming to balance the loads in each academic period. Following the description in Hnich *et al.* [2002], we can have both an integer model $M_X$ and set model $M_S$.

Given $m$ courses, and $n$ periods, $a, b$ are the minimum and maximum academic load allowed per period, $c, d$ are the minimum and maximum number of courses allowed per period, $t_i$ specifies the number of credits for course $i$, and $R$ is a set of prerequisite pairs $\langle i, j \rangle$ specifying that course $i$ must be taken before course $j$.

We introduce a set of auxiliary variables $l_j$, which is shared by both models, to represent the academic load in period $j$ as well as a variable $u$ representing the maximum academic load in any period, i.e. $u = \max\{l_j \mid 1 \leq j \leq n\}$. The objective function simply minimizes $u$. We also introduce another set of shared auxiliary variables $q_j$ to represent the number of courses assigned to a period. We have $D_{init}(u) = D_{init}(l_j) = [\,0\,..\,\Sigma_{i=1}^{m} t_i\,]$ and $D_{init}(q_j) = [\,1\,..\,m\,]$.

We have the following constraints that are common to both models:

—(B1.1) load allowed per period: $\forall 1 \le j \le n.\ \ a \le l_j \le b$

—(B1.2) number of courses allowed per period: $\forall 1 \le j \le n.\ \ c \le q_j \le d$

We also add the following redundant constraints:

—(B2.1) all the credits must be fulfilled: $(\sum_{j=1}^{n} l_j) = (\sum_{i=1}^{m} t_i)$

—(B2.2) all the courses must be taken: $(\sum_{j=1}^{n} q_j) = m$

In the integer model, $M_X$, the variables $X = \{x_i | 1 \le i \le m\}$ represent the period to which course $i$ is assigned and $D_{init}(x_i) = [\,1\,..\,n\,]$ for all $1 \le i \le m$. The constraints for the integer model $M_X$ are:

—(BX1) $l_j$ is the load taken in period $j$: $\forall 1 \le j \le n.\ \ (\sum_{i=1}^{m}((x_i = j) \times t_i)) = l_j$

—(BX2) $q_j$ is the number of courses in period $j$: $\forall 1 \le j \le n.\ \ (\sum_{i=1}^{m}(x_i = j)) = q_j$

—(BX3) courses are taken respecting prerequisites: $\forall \langle i, j \rangle \in R.\ \ x_i < x_j$

In the set model, the set variables $S = \{S_j | 1 \le j \le n\}$ represent the set of courses assigned to period $j$ and $D_{init}(S_j) = [\,\emptyset\,..\,\{1, \ldots, m\}\,]$ for all $1 \le j \le n$. The constraints for the set model $M_S$ are:

—(BS1) No course is taken twice: $\forall 1 \le i < j \le n.\ \ S_i \cap S_j = \emptyset$

—(BS2) $l_j$ is the load in period $j$: $\forall 1 \le j \le n.\ \ (\sum_{i \in S_j} t_i) = l_j$

—(BS3) $q_j$ is the number or courses in period $j$: $\forall 1 \le j \le n.\ \ |S_j| = q_j$

—(BS4) courses are taken respecting prerequisites:

$$\forall \langle i, j \rangle \in R. \forall 1 \le k \le n-1. \forall 1 \le k' \le k.\ \ (i \in S_k) \Rightarrow (j \notin S_{k'})$$

We can use the set channels to combine the two models, $x_i = j \Leftrightarrow i \in S_j$ for all $1 \le i \le m, 1 \le j \le n$   □

### 3.4 Channels between Set and Boolean Models

A very uncommon form of redundant modeling is when we give a set model and a Boolean version of this model. The reason it is uncommon is that there is no natural gain in expressiveness in moving to the Boolean model.

Suppose the set variables are $\{S_1, \ldots, S_k\}$. where $D_{init}(S_i) = [\,\emptyset\,..\,\{1, \ldots, n\}\,]$, and the Boolean variables are $z_{ij}, 1 \le i \le k, 1 \le j \le n$. The *set2bool channel function* $\backsimeq$ is defined as $\backsimeq(j \in S_i) = (z_{ij} = 1)$ and $\backsimeq(j \notin S_i) = (z_{ij} = 0)$. The *set2bool channel* $C_{\backsimeq}$ is equivalent to

$$\bigwedge_{i=1}^{k} \bigwedge_{j=1}^{n} (j \in S_i \Leftrightarrow z_{ij} = 1)$$

With $\backsimeq$ channel, we can map common set constraints ($c$) to Boolean constraints ($\backsimeq(c)$) as given in Figure 1. We shall prove that set bounds propagation of set constraints ($c$) is equivalent to domain propagation for the corresponding Boolean constraints ($\backsimeq(c)$).

| $c$ | $\rightharpoondown(c)$ |
|---|---|
| $S_i = \emptyset$ | $\{z_{ij} = 0 \mid 1 \leq j \leq k\}$ |
| $S_a \leq S_b$ | $\{z_{aj} \leq z_{bj} \mid 1 \leq j \leq k\}$ |
| $S_a \cap S_b = \emptyset$ | $\{z_{aj} + z_{bj} \leq 1 \mid 1 \leq j \leq k\}$ |
| $S_a = S_b \cup S_c$ | $\{z_{bj} \leq z_{aj} \mid 1 \leq j \leq k\} \cup \{z_{cj} \leq z_{aj} \mid 1 \leq j \leq k\} \cup$ |
| | $\{z_{aj} \leq z_{bj} + z_{cj} \mid 1 \leq j \leq k\}$ |
| $S_a = S_b \cap S_c$ | $\{z_{aj} \leq z_{bj} \mid 1 \leq j \leq k\} \cup \{z_{aj} \leq z_{cj} \mid 1 \leq j \leq k\} \cup$ |
| | $\{z_{bj} + z_{cj} \leq z_{aj} + 1 \mid 1 \leq j \leq k\}$ |
| $S_a = S_b - S_c$ | $\{z_{aj} \leq z_{bj} \mid 1 \leq j \leq k\} \cup \{z_{aj} + z_{cj} \leq 1 \mid 1 \leq j \leq k\} \cup$ |
| | $\{z_{bj} - z_{cj} \leq z_{aj} \mid 1 \leq j \leq k\}$ |
| $\|S_i\| = m$ | $\{m = \Sigma_{j=1}^{n} z_{ij}\}$ |

Fig. 1. Mapping of Common Set Constraints to Boolean Constraints

## 4. PROPAGATION REDUNDANT CONSTRAINTS

We shall be interested in reasoning about redundancy with respect to sets of propagators. We say a set of propagators $F_1$ is *stronger* than a set of propagators $F_2$, written $F_1 \gg F_2$, if $solv(F_1, D) \sqsubseteq solv(F_2, D)$ for all domains $D \sqsubseteq D_{init}$. We say a set of propagators $F_1$ is *equivalent* to a set of propagators $F_2$, written $F_1 \approx F_2$, if $solv(F_1, D) = solv(F_2, D)$ for all domains $D \sqsubseteq D_{init}$. A propagator $f$ is made *propagation redundant* by a set of propagators $F$ if $F \gg \{f\}$. Our main aim is to discover and eliminate propagation redundant constraints and/or propagators.

In redundant modeling, each model is logically redundant with respect to the other model plus the channeling constraints. In general, the propagators defined for two viewpoints act in different ways and discover information at different stages in the search. However, we show two possibilities in which propagation caused by some constraints in one model can be made redundant by: (a) propagation induced from constraints in the other model through channels and (b) propagation of the channels themselves. In order to show that, we need some useful lemmas in comparing the propagation strength of logically redundant constraints.

### 4.1 Useful Lemmas

It is clear that a constraint $c_2$ that is logically redundant with respect to constraint $c_1$ is also propagation redundant with respect to $c_1$.

LEMMA 4.1. *If* $\models D_{init} \wedge c_1 \rightarrow c_2$ *then* $\{dsb(c_1)\} \gg \{dsb(c_2)\}$.

PROOF. Follows immediately from Lemma 2.5. □

Typically though a logically redundant constraint $c_2$ is made logically redundant by a *conjunction* of other constraints. It is well known that in general the domain (and set bounds) propagation of a conjunction of constraints is *not* equivalent to applying the domain (and set bounds) propagators individually.

LEMMA 4.2. $\{dsb(c_1 \wedge c_2)\} \gg \{dsb(c_1), dsb(c_2)\}$.

PROOF. We just show the case for domain propagation of integer variables. Similar arguments apply for set bounds propagation of set variables. Suppose to the contrary that $\{dom(c_1), dom(c_2)\}$ eliminates a value $d$ from $y \in (vars(c_1 \wedge c_2) \cap \mathcal{V}_I)$ where $d \in D(y)$, but $d \in dom(c_1 \wedge c_2)(D)(y)$. By definition of propagation solver, there can be no solution $\theta$ which satisfies $c_1 \wedge c_2$ in $D$ where $\theta(y) = d$. Hence, $\{dom(c_1 \wedge c_2)\}$ must eliminate $d$ from $y$. Contradiction. □

But there is a case where propagation of a conjunction is equivalent to propagation on the individual conjuncts.

LEMMA 4.3. *Let $c_1$ and $c_2$ be two constraints sharing at most one integer variable $x \in \mathcal{V}_I$, then $\{dsb(c_1), dsb(c_2)\} \approx \{dsb(c_1 \wedge c_2)\}$.*

PROOF. We have $\{dsb(c_1 \wedge c_2)\} \gg \{dsb(c_1), dsb(c_2)\}$ by Lemma 4.2.

To show $\{dsb(c_1), dsb(c_2)\} \gg \{dsb(c_1 \wedge c_2)\}$, suppose $\theta_1$ is a solution of $c_1$ and $\theta_2$ is a solution of $c_2$, where $\theta_1(x) = \theta_2(x)$. Then if $\theta_1 \in D$ and $\theta_2 \in D$ we have that $\theta_1 \cup \theta_2 \in D$ and $\theta_1 \cup \theta_2$ is a solution of $c_1 \wedge c_2$.

Suppose $dsb(c_1 \wedge c_2)$ eliminates a value $d$ from $y \in \mathcal{V}_I$. Then there can be no solution $\theta$ of $c_1 \wedge c_2$ in $D$, where $\theta(y) = d$. Assume w.l.o.g. that $y \in vars(c_1)$. Suppose to the contrary that exists solution of $c_1$, $\theta_1 \in D$ where $\theta(y) = d$. Now if there exists a solution $\theta_2 \in D$ of $c_2$ where $\theta_2(x) = \theta_1(x)$ then we have a contradiction. Otherwise there is no such $\theta_2$ hence $dsb(c_2)(D)$ eliminates the value $\theta_1(x)$ from the domain of $x$, and hence eliminates $\theta_1$ as a solution of $c_1$.

Suppose $dsb(c_1 \wedge c_2)$ adds a value $d$ to $\inf_D(S)$ where $S \in \mathcal{V}_S$. Then there is no solution $\theta$ of $c_1 \wedge c_2$ in $D$, where $d \notin \theta(S)$. Assume w.l.o.g. that $S \in vars(c_1)$. Suppose to the contrary that exists solution of $c_1$, $\theta_1 \in D$ where $d \notin \theta_1(S)$. Now if there exists a solution $\theta_2 \in D$ of $c_2$ where $\theta_2(x) = \theta_1(x)$ then we have a contradiction. Hence there is no such $\theta_2$ thus $dsb(c_2)(D)$ eliminates the value $\theta_1(x)$ from the domain of $x$, and hence eliminates $\theta_1$ as a solution of $c_1$.

Suppose $dsb(c_1 \wedge c_2)$ eliminates a value $d$ from $\sup_D(S)$ where $S \in \mathcal{V}_S$. Then there is no solution $\theta$ of $c_1 \wedge c_2$ in $D$, where $d \in \theta(S)$. Assume w.l.o.g. that $S \in vars(c_1)$. Suppose to the contrary that exists solution of $c_1$, $\theta_1 \in D$ where $d \notin \theta_1(S)$. Now if there exists a solution $\theta_2 \in D$ of $c_2$ where $\theta_2(x) = \theta_1(x)$ then we have a contradiction. Hence there is no such $\theta_2$ $dsb(c_2)(D)$ eliminates the value $\theta_1(x)$ from the domain of $x$, and hence eliminates $\theta_1$ as a solution of $c_1$. □

Lemma 4.3 also allows us to always state that $\{dsb(c), dsb(c_x)\} \approx \{dsb(c \wedge c_x)\}$ where $c_x$ is any constraint on a single integer variable.

Note that the result above does not hold when the single variable shared is a set variable, because we only apply set bounds propagation. If we did use set domain propagators the result readily extends to the case where a single shared variable is a set variable.

EXAMPLE 4.4. Consider the set constraints $c_1 \equiv S \in \{\{1\}, \{2,3\}\}$ and $c_2 \equiv S \in \{\{2\}, \{1,3\}\}$, then $dsb(c_1)(D) = dsb(c_1)(D) = D$ where $D(S) = \{\emptyset..\{1,2,3\}\}$, while $dsb(c_1 \wedge c_2)(D)$ is a false domain since $c_1 \wedge c_2$ is unsatisfiable. Note that $dom(c_1)(D) = D_1$ where $D_1(S) = \{\{1\}, \{\{2,3\}\}$ and $dom(c_2)(D_1)(S) = \emptyset$. □

A corollary of Lemma 4.3 is that we can determine domain consistency of an entire integer CSP with tree structure just using the individual domain propagators, since we can repeatedly apply the above lemma to break the conjunction of constraints into individual constraints. This is quite related to the "backtrack-free" approach to solving CSPs with tree structure of Freuder [Freuder 1982].

## 4.2 Propagation Redundancy Through Channels

To show that the propagation caused by some constraints in one model is subsumed by propagation induced from constraints in the other model through channels, we need to break up the consideration of a constraint into individual propagation rules. The following lemma ensures that the domain and set bounds propagator of a constraint is equivalent to the union of the propagation rules implemented by the propagator.

LEMMA 4.5. *Given a constraint $c$. Then $\{dsb(c)\} \approx prop(dsb(c))$.*

PROOF. We have $\{dsb(c)\} \gg prop(dsb(c))$ by Lemma 2.5 and and Lemma 4.1.
It remains to show that $prop(dsb(c)) \gg \{dsb(c)\}$. Suppose $D' = dsb(c)(D)$, and $D'' = solv(prop(dsb(c)), D)$, We have the following cases:

—Suppose $d \notin dom(c)(D)(x_i)$ and $d \in D(x_i)$ where $x_i \in \mathcal{V}_I$. Then clearly $dsb(c)$ implements a rule $r \equiv \overline{D} \rightarrowtail x_i \neq d$ where $\overline{D} = \wedge_{j=1}^{n} \wedge_{d' \in D_{init}(x_j) - D(x_j)} x_j \neq d'$. Now since $r \in rules(dsb(c))$ we have that $prop(dsb(c)) \gg \{r\}$ by the definition of $prop(dsb(c))$. Hence $d \notin D''(x_i)$.

—Suppose $d \in \inf_{D'}(S_i)$ and $d \notin \inf_D(S_i)$ where $S_i \in \mathcal{V}_S$. Clearly $dsb(c)$ implements a rule $r \equiv \overline{D} \rightarrowtail d \in S_i$ where

$$\overline{D} = \wedge_{j=1}^{n} \wedge_{d' \in \inf_D(S_j) - \inf_{D_{init}}(S_j)} d' \in S_j$$
$$\wedge_{k=1}^{n} \wedge_{d'' \in \sup_{D_{init}}(S_k) - \sup_D(S_k)} d'' \notin S_k$$

As above since $r \in rules(dsb(c))$ we have that $prop(dsb(c)) \gg \{r\}$ by the definition of $prop(dsb(c))$. Hence $d \in \inf_{D'}(S_i)$.

—$d \notin \sup_{D'}(S_i)$ and $d \in \sup_D(S_i)$ where $S_i \in \mathcal{V}_S$. This case is similar to the previous case.

Since $D''$ has all the modifications to the domain made by $D'$ we have that $D'' \sqsubseteq D'$ and $prop(dsb(c)) \gg \{dsb(c)\}$. □

A propagation rule $C_1 \rightarrowtail c_1$ *directly subsumes* a rule $C_2 \rightarrowtail c_2$ if $\models (D_{init} \wedge C_2) \rightarrow C_1$ and $\models (D_{init} \wedge c_1) \rightarrow c_2$. A propagation rule that is mapped by a channel function to a rule directly subsumed by another propagation rule is propagation redundant. We extend channel functions to map conjunctions of atomic constraints in the obvious manner $\Diamond(c_1 \wedge \cdots \wedge c_n) = \Diamond(c_1) \wedge \cdots \wedge \Diamond(c_n)$.

LEMMA 4.6. *Let $C \rightarrowtail c$ be a propagation rule on $Y$ variables, and $C' \rightarrowtail c'$ be a propagation rule on $X$ variables. If*

$$\models D_{init} \wedge \Diamond^{-1}(C) \rightarrow C' \text{ and } \models D_{init} \wedge c' \rightarrow \Diamond^{-1}(c),$$

*then*

$$\{C' \rightarrowtail c'\} \cup F_\Diamond \gg \{C \rightarrowtail c\}.$$

PROOF. Suppose $C \rightarrowtail c$ is applied to enforce $c$. Then $\models D \rightarrow C$. Now if $D_1 = solv(F_\Diamond, D)$ then by the definition of $F_\Diamond$ clearly $\models D_1 \rightarrow \Diamond^{-1}(C)$ and hence $\models D_1 \rightarrow C'$. By the definition of the propagation rule $C' \rightarrowtail c'$ we have $\models D_2 \rightarrow c'$, where $D_2 = (C' \rightarrowtail c')(D_1)$; hence $\models D_2 \rightarrow \Diamond^{-1}(c)$. Then, by the definition of $F_\Diamond$, $\models D_3 \rightarrow c$, where $D_3 = solv(F_\Diamond, D_2)$. □

We can straightforwardly lift the above results to talk about propagation rules that are directly subsumed by the domain and set bounds propagator for a constraint, and then lift to a set of propagation rules implemented by some propagator.

THEOREM 4.7. *Let $C \rightarrowtail c$ be a propagation rule on $Y$ variables, and $c_X$ be a constraint on $X$ variables. If*

$$\models (D_{init} \wedge c_X \wedge \Diamond^{-1}(C)) \rightarrow \Diamond^{-1}(c),$$

*then*

$$\{dsb(c_X)\} \cup F_\Diamond \gg \{C \rightarrowtail c\}.$$

PROOF. By Lemma 4.5 we have that $dsb(c_X)$ implements the propagation rule $\Diamond^{-1}(C) \rightarrow \Diamond^{-1}(c)$. Hence by Lemma 4.6 the result holds. □

COROLLARY 4.8. *Let $c_Y$ be a constraint on $Y$ variables, and $c_X$ be a constraint on $X$ variables. If*

$$\models (D_{init} \wedge c_X \wedge \Diamond^{-1}(C)) \rightarrow \Diamond^{-1}(c) \text{ for all } C \rightarrowtail c \in prop(dsb(c_Y)),$$

*then*

$$\{dsb(c_X)\} \cup F_\Diamond \gg \{dsb(c_Y)\}.$$

EXAMPLE 4.9. Consider the (LY2.1) constraints $c_Y \equiv y_j = 3i - 2 \Leftrightarrow y_{j+(i+1)} = 3i - 1$ of the Langford's Problem (Example 3.1), where $1 \leq i \leq 9$ and $1 \leq j \leq 27 - 2(i+1)$. The propagation rules for $dsb(c_Y)$ are

$$y_j = 3i - 2 \rightarrowtail y_{j+(i+1)} = 3i - 1$$
$$y_{j+(i+1)} = 3i - 1 \rightarrowtail y_j = 3i - 2$$
$$y_j \neq 3i - 2 \rightarrowtail y_{j+(i+1)} \neq 3i - 1$$
$$y_{j+(i+1)} \neq 3i - 1 \rightarrowtail y_j \neq 3i - 2$$

We have that for $c_X \equiv x_{3i-1} = x_{3i-2} + (i+1)$ of (LX2.1), $\models D_{init} \wedge c_X \wedge \bowtie (C) \rightarrow \bowtie(c)$ for all propagation rules above. For example the first rule is mapped to $x_{3i-2} = j \rightarrowtail x_{3i-1} = j + i + 1$. Hence the conditions of Corollary 4.8 hold and $dsb(c_Y)$ is propagation redundant.

We can similarly show the (LY2.2) constraints $c'_Y \equiv y_j = 3i - 2 \Leftrightarrow y_{j+2(i+1)} = 3i$, where $1 \leq i \leq 9$ and $1 \leq j \leq 27 - 2(i+1)$, are propagation redundant using $c_X \wedge c'_X$, where $c'_X \equiv x_{3i} = x_{3i-1} + (i+1)$ of (LX2.2). Although model $M_X$ does not include a domain propagator for $c_X \wedge c'_X$, we can still show propagation redundancy since $\{dsb(c_X), dsb(c'_X)\} \approx \{dsb(c_X \wedge c'_X)\}$ by Theorem 4.3. Similar reasoning applies to show that the (LY3) constraints $y_j \neq 3i - 2$, where $1 \leq i \leq 9$ and $(28 - 2(i+1)) \leq j \leq 27$, are made propagation redundant by $c_X \wedge c'_X$. □

For brevity we shall introduce pseudo atomic constraints $x \leq d$ equivalent to the conjunction $x \neq d+1, \ldots, x \neq \sup_{D_{init}}(x)$ and $x \geq d$ equivalent to the conjunction $x \neq \inf_{D_{init}}(x), \ldots, x \neq d-1$, to discuss the next example.

EXAMPLE 4.10. Consider the (BX2) constraints $c_X \equiv (\sum_{i=1}^{m}(x_i = j)) = q_j$ of the balanced academic curriculum problem (Example 3.5), where $1 \leq j \leq n$. The propagation rules $C \rightarrowtail c$ for $dsb(c_X)$ are

$$q_j \leq d \wedge x_{i_1} = j \wedge \cdots \wedge x_{i_d} = j \rightarrowtail x_i \neq j$$
$$x_{i_1} = j \wedge \cdots \wedge x_{i_d} = j \rightarrowtail q_j \geq d$$

for all $I = \{i_1, \ldots, i_d\} \subseteq \{1, \ldots, m\}$ and $i \in \{1, \ldots, m\} - I$; and

$$q_j \geq d \wedge x_{i_1} \neq j \wedge \cdots \wedge x_{i_{m-d}} \neq j \rightarrowtail x_i = j$$
$$x_{i_1} \neq j \wedge \cdots \wedge x_{i_{m-d}} \neq j \rightarrowtail q_j \leq d$$

for all $I = \{i_1, \ldots, i_{m-d}\} \subseteq \{1, \ldots, m\}$ and $i \in \{1, \ldots, m\} - I$. Notice that all the atomic constraints involving $q_j$ are mapped to themselves by $\{\}$, since $q_j$ is shared by the two models. The rules are mapped to

$$q_j \leq d \wedge i_1 \in S_j \wedge \cdots \wedge i_d \in S_j \rightarrowtail i \notin S_j$$
$$i_1 \in S_j \wedge \cdots \wedge i_d \in S_j \rightarrowtail q_j \leq d$$
$$q_j \geq d \wedge i_1 \notin S_j \wedge \cdots \wedge i_{m-d} \notin S_j \rightarrowtail i \in S_j$$
$$i_1 \notin S_j \wedge \cdots \wedge i_{m-d} \notin S_j \rightarrowtail q_j \geq d$$

We have that for $c_S \equiv |S_j| = q_j$ of (BS3), $\models (D_{init} \wedge c_S \wedge \{\}(C)) \rightarrow \{\}(c)$ for all the propagation rules above. Hence, $dsb(c_X)$ is propagation redundant using Corollary 4.8.

Similar reasoning applies to show that each constraint $(\sum_{i=1}^{m}((x_i = j) \times t_i)) = l_j$ of (BX1) is made propagation redundant by $(\sum_{i \in S_j} t_i) = l_j$ of (BS2), where $1 \leq j \leq n$. □

Often a single constraint does not capture all the propagation effects of a constraint on the other side of the permutation model. In that case we may need to find for each particular propagation rule, a constraint on the other side that causes the same propagation to occur.

THEOREM 4.11. *Let $c_Y$ be a propagator on $Y$ variables. Suppose for each $r \equiv (C \rightarrowtail c) \in prop(dsb(c_Y))$, there exists constraint $imp(r)$ on $X$ variables where $\models (D_{init} \wedge imp(r) \wedge \Diamond^{-1}(C)) \rightarrow \Diamond^{-1}(c)$, then $\cup_{r \in prop(dsb(c_Y))}\{dsb(imp(r))\} \cup F_\Diamond \gg \{dsb(c_Y)\}$.*

PROOF. The proof follows straightforwardly from Lemma 4.5 and Theorem 4.7. □

EXAMPLE 4.12. Consider the (IY2.1) constraints $c_Y \equiv (y_i - y_j = 1) \Rightarrow (v_{j-i} = y_j)$ of the all intervals series problem (Example 3.2), where $1 \leq i < j \leq n$. The propagation rules for $dsb(c_Y)$ have the forms

$$r1 : y_i = k+1 \wedge y_j = k \rightarrowtail v_{j-i} = k$$
$$r2 : v_{j-i} \neq k \wedge y_j = k \rightarrowtail y_i \neq k+1$$
$$r3 : \quad y_i = k+1 \wedge I \rightarrowtail y_j \neq k$$

where in $r3$, $I$ is any conjunction of disequations on $v_{j-i}$ and $y_j$, not including $y_j \neq k$ ensuring that $v_{j-i} \neq y_j$. We can show for $imp(r1) \equiv imp(r2) \equiv imp(r3) \equiv (u_k = |x_k - x_{k+1}|)$ of (IX2) that $\models (D_{init} \wedge imp(r1) \wedge x_{k+1} = i \wedge x_k = j) \rightarrow (u_k = j - i)$ and $\models (D_{init} \wedge imp(r2) \wedge u_k \neq j - i \wedge x_k = j) \rightarrow (x_{k+1} \neq i)$. For the remaining propagation rules (r3), it is clear that $I$ must contain $v_{j-i} \neq k$ since it does not contain $y_j \neq k$ and it must force the two to be different. We can show that

$$\models (D_{init} \wedge imp(r3) \wedge u_k \neq j - i \wedge x_{k+1} = i) \rightarrow (x_k \neq j).$$

Note that even though each example propagation rule r1–r3 is made propagation redundant by the same constraint $u_k = |x_k - x_{k+1}|$ we require a different constraint for each different value of $k$.

Hence the constraint is propagation redundant by Theorem 4.11. Similarly for the other (IY2.2) constraints $(y_j - y_i = 1) \Rightarrow (v_{j-i} = y_i)$, where $1 \leq i < j \leq n$. Clearly, the redundant constraint (IY3) $(|y_1 - y_n| = 1) \wedge (v_{n-1} = \min(y_1, y_n))$ is non-propagation redundant. $\square$

### 4.3 Propagation Redundancy Caused by Channels

The channels themselves may actually restrict the possible solutions in one or both models involved. We concentrate on the $X$ model, since the restrictions on the $Y$ model can be seen by examining the inverse channel function.

A channel function $\diamondsuit$ is *restrictive* (on the variables $X$) if $\not\models D_{init} \rightarrow \exists Y C_\diamondsuit$, that is not all valuations on $X$ variables are extensible to solutions of $C_\diamondsuit$.

EXAMPLE 4.13. The permutation channel function $\bowtie$ is restrictive, for example $\{x_1 = 2, x_2 = 2\}$ cannot be extended to be a solution of $C_\bowtie$, since it requires $y_2$ to take both values 1 and 2. The Boolean channel function $\triangle$ is unrestrictive. Any valuation on $X$ variables can be extended to a solution of $C_\triangle$. However $\triangle^{-1}$ is restrictive, for example $\{z_{11} = 1, z_{12} = 1\}$ cannot be extended to a solution of $C_\triangle$ since it requires $x_1$ to be both 1 and 2. Similarly the set channel function $\{\}$ is unrestrictive while $\{\}^{-1}$ is restrictive. For example $S_1 = \{1\}, S_2 = \{1\}$ cannot be extended to a solution of $C_{\{\}}$ since it requires $x_1$ to be both 1 and 2. The set2bool channel $\simeq$ is unrestrictive in both directions. $\square$

4.3.1 *Restrictive Channel Functions.* Restrictive channel functions can themselves make constraints propagation redundant.

Smith [2000] first observe that the permutation channel makes each of the disequations between variables in either model propagation redundant. Walsh [2001] proves this holds for other notions of consistency.

LEMMA 4.14 WALSH [2001]. $F_\bowtie \gg \{dsb(x_i \neq x_j)\}$ for all $1 \leq i < j \leq n$. $\square$

EXAMPLE 4.15. Lemma 4.14 makes the (LX1) and (LY1) constraints of the Langford's Problem (Example 3.1), and the (IX1.1), (IX1.2), (IY1.1) and (IY1.2) constraints of the all intervals series (Example 3.2) propagation redundant. $\square$

Implicit in the Boolean channel is that each integer variable can take only one, and must take one value. This is represented in the Boolean model as the constraint $\sum_{j=0}^{k} z_{ij} = 1$. It is enforced by the restrictive channel function $\triangle^{-1}$.

LEMMA 4.16. $F_\triangle \gg \{dsb(\sum_{j=1}^{k} z_{ij} = 1)\}$ for all $1 \leq i \leq n$.

PROOF. The propagation rules for $dsb(\sum_{j=1}^{k} z_{ij} = 1)$ are

$$z_{ij} = 1 \rightarrowtail z_{ij'} = 0, j \neq j'$$
$$z_{i1} = 0, \ldots, z_{i(j-1)} = 0, z_{i(j+1)} = 0, \ldots, z_{ik} = 0 \rightarrowtail z_{ij} = 1$$

For the first rule, $D(z_{ij}) = \{1\}$ means that $dsb(\sum_{j=1}^{k} z_{ij} = 1)(D)(z_{ij'}) \subseteq \{0\}$. Let $D = solv(F_\triangle, D)$. Then $D(x_i) = \{j\}$ using $x_i = j \Leftrightarrow z_{ij} = 1$ and $1 \notin D(z_{ij'}), j \neq j'$ using $x_i = j' \Leftrightarrow z_{ij'} = 1$.

For the second rule, $D(z_{ij}) = \{0\}, \forall 1 \leq j' \neq j \leq k$ and $dsb(\sum_{j=1}^{k} z_{ij} = 1)(D)(z_{ij}) \subseteq \{1\}$. Let $D = solv(F_\triangle, D)$. Now $D(x_i) \cap \{1, \ldots, j-1, j+1, \ldots, k\} = \emptyset$

using $x_i = j' \Leftrightarrow z_{ij'} = 1$. Hence $D(x_i) = \{j\}$ and $0 \notin D(z_{ij})$ using $x_i = j \Leftrightarrow z_{ij} = 1$. □

EXAMPLE 4.17. The (QZ1) constraints of the $n$-Queens Problem (Example 3.3) are propagation redundant using Lemma 4.16. □

The channel function $\{\}^{-1}$ is restrictive, since each variable $x_i \in X$ can only take a single value $j$. It means that $S_j \cap S_{j'} = \emptyset$ for all $0 \leq j < j' \leq m$. It is clear that $F_{\{\}}$ makes these constraints propagation redundant.

LEMMA 4.18. $F_{\{\}} \gg \{dsb(S_j \cap S_{j'} = \emptyset)\}$ for all $1 \leq j < j' \leq m$.

PROOF. The propagation rules for $dsb(S_j \cap S_{j'} = \emptyset)$ are

$$i \in S_j \rightarrowtail i \notin S_{j'}$$
$$i \in S_{j'} \rightarrowtail i \notin S_j$$

For the first rule, assume w.l.o.g. that $dsb(S_j \cap S_{j'} = \emptyset)$ makes $i \notin S_{j'}$, then $i \in \inf_D(S_j)$. Clearly, $D' = solv(F_{\{\}}, D)$ is such that $D'(x_i) = \{j\}$ using $x_i = j \Leftrightarrow i \in S_j$. Then $i \notin \sup_{D'}(S'_j)$, using $x_i = j' \Leftrightarrow i \in S_{j'}$. A similar arguments applies to the second rule. □

EXAMPLE 4.19. Lemma 4.18 makes both the (GS1) constraints of the social golfers problem (Example 3.4) and (BS1) constraints of the balanced academic curriculum problem (Example 3.5) propagation redundant. □

4.3.2 *Unrestrictive Channel Functions.* Unrestrictive channel functions do not make any constraints (on $X$) propagation redundant. Interestingly in this case we can argue about propagation redundancy simply in terms of logical consequence.

THEOREM 4.20. *Let $\Diamond$ be an unrestrictive channel function, let $c_Y$ be a constraint on $Y$ variables, and $c_X$ a constraint on $X$ variables. If*

$$\models (D_{init} \wedge c_X \wedge C_\Diamond) \rightarrow c_Y,$$

then

$$\{dsb(c_X)\} \cup F_\Diamond \gg \{dsb(c_Y)\}.$$

PROOF. Let $C \rightarrowtail c \in prop(dsb(c_Y))$ then clearly $\models D_{init} \wedge c_X \wedge C_\Diamond \rightarrow (C \rightarrow c)$, since $\models D_{init} \wedge c_Y \rightarrow (C \rightarrow c)$ by Lemma 2.5. We show that

$$\models D_{init} \wedge c_X \rightarrow (\Diamond^{-1}(C) \rightarrow \Diamond^{-1}(c))$$

by contradiction.

Assume to the contrary, there is a solution $\theta_X$ of $D_{init} \wedge c_X \wedge \Diamond^{-1}(C) \wedge \neg \Diamond^{-1}(c)$. Now since $\theta_X$ is a solution of $D_{init} \wedge c_X$ and $\Diamond$ is non-constraining it is extensible to a solution $\theta = \theta_X \cup \theta_Y$ of $C_\Diamond$.

Now $\theta_Y$ is a solution of $C$ by the definition of $C_\Diamond$, but similarly since $\theta_X$ is not a solution of $\Diamond^{-1}(c)$, then $\theta_Y$ is not a solution of $c$. Contradiction. □

The reason the channel function must be unrestrictive for this result to hold is that the $\models (D_{init} \wedge c_X \wedge C_\Diamond) \rightarrow c_Y$ is too weak a condition in the general case.

EXAMPLE 4.21. The permutation channel function is restrictive. Now $\models C \to y_3 = 3$, where $C \equiv x_1 + x_2 < 4 \wedge C_{\bowtie}$, since the only solutions of $C$ are $\{x_1 \mapsto 1, x_2 \mapsto 2, x_3 \mapsto 3, y_1 \mapsto 1, y_2 \mapsto 2, y_3 \mapsto 3\}$ and $\{x_1 \mapsto 2, x_2 \mapsto 1, x_3 \mapsto 3, y_1 \mapsto 2, y_2 \mapsto 1, y_3 \mapsto 3\}$. But clearly it is not the case that $x_1 + x_2 < 4 \to x_3 = 3$. The problem is that the channel $C_{\bowtie}$ removes solutions of $x_1 + x_2 < 4$ like $\{x_1 \mapsto 1, x_2 \mapsto 1, x_3 \mapsto 1\}$ from consideration. □

We can use Theorem 4.20 to prove propagation redundancy of many of the propagators in our examples.

EXAMPLE 4.22. Consider the (QZ3.1) constraint $c_Z \equiv \sum_{i=1}^{n} z_{ii} \leq 1$ of the $n$-Queens Problem (Example 3.3). We can show for

$$c_X \equiv x_1 \neq x_i - i + 1 \wedge \cdots \wedge x_{i-1} \neq x_i - 1 \wedge x_{i+1} \neq x_i + 1 \wedge \cdots \wedge x_n \neq x_i + n - i$$

of (QX2.1) and (QX2.2) that $\models D_{init} \wedge c_X \wedge C_\triangle \to c_Z$. Now

$$\{dsb(c_X)\} \approx \{dsb(x_1 \neq x_i - i + 1), \ldots, dsb(x_n \neq x_i + n - i)\}$$

by Lemma 4.3. Since $\triangle$ is an unrestrictive channel function, by Theorem 4.20 we have that $dsb(c_Z)$ is propagation redundant. A similar argument applies to the (QZ3.2), (QZ4.1), (QZ4.2), (QZ4.3) and (QZ4.4) constraints.

Note that the (QZ2) constraints $\sum_{i=1}^{n} z_{ij} = 1$, where $1 \leq j \leq n$, are not propagation redundant, although the constraint $\sum_{i=1}^{n} z_{ij} \leq 1$ is made redundant by (QX1) (using a similar argument to the (QZ3.1) constraints). □

EXAMPLE 4.23. For the social golfers problem (Example 3.4), the (GS2) constraints $c_{S_1} \equiv |S_{ik}| = s$ where $1 \leq i \leq g$ and $1 \leq k \leq w$, we can show for $c_{X_1} \equiv \Sigma_{l=1}^{n}(x_{lk} = i) = s$ of (GX1) that $\models D_{init} \wedge c_{X_1} \wedge C_{\{\}} \to c_{S_1}$. For the (GS3) constraints $c_{S_2} \equiv |S_{i_1 k_1} \cap S_{i_2 k_2}| \leq 1$ where $1 \leq i_1 \neq i_2 \leq g$ and $1 \leq k_1 \neq k_2 \leq w$, we can also show for $c_{X_2} \equiv \neg(x_{l_1 k_1} = x_{l_2 k_1} \wedge x_{l_1 k_2} = x_{l_2 k_2})$ of (GX2) that $\models D_{init} \wedge c_{X_2} \wedge C_{\{\}} \to c_{S_2}$. Since $\{\}$ is an unrestrictive channel function by Theorem 4.20 we have both $dsb(c_{S_1})$ and $dsb(c_{S_2})$ are propagation redundant. □

EXAMPLE 4.24. For the balanced academic curriculum problem (Example 3.5), the (BS4) constraints $c_S \equiv (i \in S_k) \Rightarrow (j \notin S_{k'})$ where $\langle i, j \rangle \in R$, $1 \leq k \leq n-1$, and $1 \leq k' \leq k$, we can show for $c_X \equiv x_i < x_j$ of (BX3) that $\models (D_{init} \wedge c_X \wedge C_{\{\}}) \to c_S$. Since $\{\}$ is an unrestrictive channel function by Theorem 4.20 we have that $dsb(c_S)$ is propagation redundant. □

In part because the $\backsimeq$ channel is unrestrictive in both directions, we can prove that set bounds propagation does not provide any more propagation strength than the mapping of set constraints to Booleans. First, we introduce the notion of *nogood constraints*.

A *nogood* constraint $c$ is one where every valuation over $vars(c)$ in $D_{init}$ except one valuation $\theta$ is a solution of $c$. We call the non-solution valuation $\theta$ the nogood of $c$. The following lemma identify the condition where propagation of the conjunction of a nogood constraint $c_1$ and a constraint $c_2$ of integer variables such that $vars(c_2) \subseteq vars(c_1)$ is equivalent to propagation on the individual conjuncts. The condition requires that each valuation $\theta' \in D_{init}$ differing from the nogood $\theta$ by only one assignment must be a solution of $c_2$.

LEMMA 4.25. *Let $c_1$ be a nogood constraint with $vars(c_1) = \{x_1, \ldots, x_n\} \subseteq \mathcal{V}_I$ and the nogood $\theta$, and $c_2$ be a constraint with $vars(c_2) \subseteq vars(c_1)$. If for all valuations $\theta' \in D_{init}$, such that there exists $1 \leq j \leq n$ and $\theta'(x_i) = \theta(x_i)$ for all $1 \leq i \neq j \leq n$, are solutions of $c_2$, then $\{dom(c_1), dom(c_2)\} \approx \{dom(c_1 \wedge c_2)\}$.*

PROOF. By Lemma 4.2 we have that $\{dom(c_1 \wedge c_2)\} \gg \{dom(c_1), dom(c_2)\}$.

It remains to show $\{dom(c_1), dom(c_2)\} \gg \{dom(c_1 \wedge c_2)\}$. Suppose to the contrary that $d \notin dom(c_1 \wedge c_2)(D)(x_k)$ and $d \in solv(\{dom(c_1), dom(c_2)\}, D)(x_k)$ where $1 \leq k \leq n$ and $d \in D(x_k)$.

First, we show that the nogood $\theta \in D$. Otherwise, when $\theta \notin D$, there must exists $1 \leq j \leq n$ such that $\theta(x_j) \notin D(x_j)$. Then any valuation $\theta'' \in D$ on $vars(c_1)$ which makes $c_2$ true is also a solution of $c_1$ since $\theta''(x_j) \neq \theta(x_j)$. Thus $dom(c_2)$ removes any values removed by $dom(c_1 \wedge c_2)$, Contradiction.

Second, we show that $\theta(x_k) = d$. Otherwise, when $\theta(x_k) \neq d$, any solution $\theta'' \in D$ on $vars(c_1)$ which makes $c_2$ true is also a solution of $c_1$. Hence $d \notin dom(c_2)(D)(x_k)$. Contradiction.

Now, since $d \in dom(c_1)(D)(x_k)$, it is not the case that $D(x_i) = \{\theta(x_i)\}$, for $1 \leq k \neq i \leq n$. Thus there exists $j \neq k$ and $x_j \in vars(c_2)$ such that $|D(x_j)| \geq 2$. Then we have $d_j \in D(x_j)$ such that $d_j \neq \theta(x_j)$.

Consider the valuation $\theta'$ defined as $\theta'(x_i) = \theta(x_i), 1 \leq i \neq j \leq n$ and $\theta'(x_j) = d_j$, note that $\theta'(x_k) = \theta(x_k) = d$. By construction $\theta' \in D$ is a solution of $c_1$ and also of $c_2$ by the conditions of the lemma. Hence, $d \in dom(c_1 \wedge c_2)(D)(x_k)$ by the definition of $dom$. Contradiction. □

The following theorem prove that set bounds propagation of set constraints is equivalent to domain propagation of the corresponding Boolean constraints. It does however still provide a more efficient implementation.

THEOREM 4.26. *Let $dsb(c)$ be the set bounds propagator for set constraint $c$, let $\simeq(c)$ be the Boolean equivalent of $c$, then (a) $\{dsb(c)\} \cup F_\simeq \gg \{dsb(c') \mid c' \in \simeq(c)\}$ and (b) $\{dsb(c') \mid c' \in \simeq(c)\} \cup F_\simeq \gg \{dsb(c)\}$.*

PROOF. We can verify that $\models D_{init} \wedge C_\simeq \rightarrow (c \leftrightarrow (\wedge\{c' \mid c' \in \simeq(c)\}))$ for each of the constraints $c$ above. Since $\simeq$ is an unrestrictive channel this immediately gives us that $\{dsb(c)\} \cup F_\simeq \gg \{dsb(c') \mid c' \in \simeq(c)\}$ using Theorem 4.20 and Lemma 4.1.

Now since $\simeq^{-1}$ is also an unrestrictive channel we also have by Theorem 4.20 that $\{dsb(\wedge\{c' \mid c' \in \simeq(c)\})\} \cup F_\simeq \gg \{dsb(c)\}$. It remains to show that $\{dsb(\wedge\{c' \mid c' \in \simeq(c)\})\} \approx \{dsb(c') \mid c' \in \simeq(c)\}$.

For $c$ of the form $S_i = \emptyset$, $S_a \subseteq S_b$, $S_a \cap S_b = \emptyset$, and $|S_i| = m$ no two constraints in $\simeq(c)$ share a variable. Hence the result holds by Lemma 4.3.

For the remaining constraints we can, again by Lemma 4.3 simply consider the 3 Boolean constraints for a particular $j$, since the remaining constraints do not share variables. We show the case for $c \equiv S_a = S_b \cup S_c$, the other constraints $S_a = S_b \cap S_c$ and $S_a = S_b - S_c$ are similar. We first show that

$$\{dsb(z_{aj} \leq z_{bj} + z_{cj} \wedge (z_{bj} \leq z_{aj} \wedge z_{cj} \leq z_{aj}))\}$$
$$\approx \{dsb(z_{aj} \leq z_{bj} + z_{cj}), dsb(z_{bj} \leq z_{aj} \wedge z_{cj} \leq z_{aj})\}$$

Now $c_1 \equiv z_{aj} \leq z_{bj} + z_{cj}$ is a nogood constraint with nogood $\{z_{aj} \mapsto 1, z_{bj} \mapsto 0, z_{cj} \mapsto 0\}$, while the valuations $\{z_{aj} \mapsto 0, z_{bj} \mapsto 0, z_{cj} \mapsto 0\}$, $\{z_{aj} \mapsto 1, z_{bj} \mapsto$ $1, z_{cj} \mapsto 0\}, \{z_{aj} \mapsto 1, z_{bj} \mapsto 0, z_{cj} \mapsto 1\}$ are all solutions of $c_2 \wedge c_3$ where $c_2 \equiv z_{bj} \leq z_{aj}$ and $c_3 \equiv \wedge z_{cj} \leq z_{aj}$. Hence by Lemma 4.25 we have that $\{dsb(c_1 \wedge (c_2 \wedge c_3))\} \approx \{dsb(c_1), dsb(c_2 \wedge c_3)\}$.

Now $c_2$ and $c_3$ share only one variable, hence $\{dsb(c_2 \wedge c_3)\} \approx \{dsb(c_2), dsb(c_3)\}$ by Lemma 4.3. Hence $\{dsb(c_1 \wedge (c_2 \wedge c_3))\} \approx \{dsb(c_1), dsb(c_2), dsb(c_3)\}$ as desired. □

## 5. DISCUSSIONS

Propagation redundancy reasoning is highly sensitive to the exact form of the constraints in the model. A slightly different model may have different degree of propagation redundancy.

EXAMPLE 5.1. Another version of the second model of the all intervals series constraints, replaces (IY2.1) $(y_i - y_j = 1) \Rightarrow (v_{j-i} = y_j)$ and (IY2.2) $(y_j - y_i = 1) \Rightarrow (v_{j-i} = y_i)$ by the (IY4) constraints

$$(|y_i - y_j| = 1) \Rightarrow (v_{j-i} = \min(y_i, y_j)),$$

where $1 \leq i < j \leq n$. This combined constraint is not propagation redundant. There are cases where it can propagate information not discovered by the first model. For example $c_Y \equiv (|y_3 - y_7| = 1) \Rightarrow (v_4 = \min(y_3, y_7))$ when $D(y_3) = \{5\}$ and $D(y_7) = \{4, 6\}$, makes $dsb(c_Y)(D)(v_4) = \{4, 5\}$, but there is no equivalent propagation in the $X$ model. □

Our definition of set bounds propagators agree with the definitions of the original set bounds propagation system [Gervet 1997]. There are stronger set based propagators (see [Azevedo and Barahona 2000; Müller 2001]) that reason more about cardinalities.

EXAMPLE 5.2. Suppose we have constraints $S_1 \subseteq S_2, |S_1| = 2, |S_2| = 1$ where $S_1$ and $S_2$ range over $\{1, 2, 3\}$. The propagators $\{dsb(S_1 \subseteq S_2)\} \cup \{dsb(|S_1| = 2)\} \cup \{dsb(|S_2| = 1)\}$ lead to no propagation. But if we propagate cardinality information, we immediately get failure. □

We can model *most* common cardinality reasoning using additional propagators. Let every set variable $S$ has an attached integer variable $x_S$ representing its cardinality. This attachment comes with implicit cardinality propagator: $dsb(|S| = x_S)$. We can then understand stronger cardinality reasoning as additional constraints using these implicit cardinality variables.

EXAMPLE 5.3. With cardinality reasoning, the set constraints $c \equiv S \subseteq S'$ defines implicitly the constraint: $x_S \leq x_{S'}$ where $|S| = x_S$ and $|S'| = x_{S'}$. Hence, the correct propagators for $c$ are $\{dsb(S \subseteq S')\} \cup \{dsb(x_S \leq x_{S'})\}$, together with the implicit propagators, $\{dsb(|S| = x_S)\} \cup \{dsb(|S'| = x_{S'})\}$. □

With cardinality reasoning, for a set constraint to be propagation redundant, we need to prove the redundancy of the extra cardinality propagators too. For the examples in this paper, this is straightforward. However, some cardinality reasoning is difficult to model using additional propagators as the reasoning relies on dynamic information during propagation.

EXAMPLE 5.4. The set constraint $x = \sum_{i \in S} u_i$ where $U = \{u_1, \ldots, u_n\}$ and $D_{init}(S) = \{\emptyset \ldots \{1, \ldots, n\}\}$ is supported in ILOG Solver 4.4. The cardinality reasoning can be computed efficiently by:

$$x \geq (\sum_{i \in \inf_D(S)} u_i) + (\min\{u_j | j \in \sup_D(S)\} * (|S| - |\inf_D(S)|))$$

and

$$x \leq (\sum_{i \in \inf_D(S)} u_i) + (\max\{u_j | j \in \sup_D(S)\} * (|S| - |\inf_D(S)|)),$$

where $D$ is the current domain during propagation. Such reasoning is rather difficult to model as separate propagators. □

## 6. EXPERIMENTS

We can take advantage of the reasoning about propagation redundancy to eliminate propagators that are propagation redundant. We then get a model with exactly the same propagation strength but with less propagators. This can translate into *faster* propagation.[4]

In the following experiments, all the benchmarks are executed using ILOG Solver 4.4 on Sun Ultra 5/400 workstations running Solaris 8. The first column of each table describes the models under comparison. In the case of combined models, we have the choice of labeling the variables from just one model, or from both models together. The second column indicates the choices of search variables. However, the question of choosing the "best" set of search variables that gives the smallest search space is out of the scope of this paper. The third column and beyond give the problem instances. For each problem instance, we measure the total number of fails and CPU time (in seconds) to compare the performance of the different models. In other words, each row shows the performance of a particular model across the different problem instances. Table entries marked with a "—" mean failure to solve the problem after one hour of execution. Table entries in bold letters highlight the model with the smallest number of fails and best runtime for each problem instance.

### 6.1 Langford's Problem

Table I compares the different models for finding all the solutions of the Langford's Problem. Variable ordering is based on the first-fail principle and value ordering is from the least to the greatest. The models under comparison include the single models: $M_X$ and $M_Y$, the *full* combined model $M_X + C_{\bowtie} + M_Y$, and an *opt*imized combined model LX2.1 + LX2.2 + $C_{\bowtie}$ as discussed in Examples 4.9 and 4.15. The *opt* model corresponds to the *minimal combined model* of Smith [2000]. Our results agree with those presented by Smith, where the *opt* models are faster and maintain the same number of fails as the *full* model for all three set of search variables. The *opt* model with search variables $X \cup Y$ is the fastest among all the models under comparison.

---

[4]Note there is no guarantee since e.g. the number of propagation steps may have increased.

Table I. Results of the Langford's Problem

| Model | Search Variables | $(3 \times 10)$ fails (sec) | $(3 \times 11)$ fails (sec) | $(4 \times 14)$ fails (sec) | $(4 \times 15)$ fails (sec) |
|---|---|---|---|---|---|
| $M_X$ | $X$ | 3114 (1.33) | 14512 (6.09) | 83068 (87.21) | 351126 (430.54) |
| full | $X$ | 1318 (13.66) | 5177 (62.5) | 20885 (1499.09) | 78556 (6264.74) |
| opt | $X$ | 1318 (1.21) | 5177 (4.77) | 20885 (51.1) | 78556 (183.38) |
| $M_Y$ | $Y$ | — — | — — | — — | — — |
| full | $Y$ | 1059 (11.17) | 3958 (50.67) | 8139 (707.77) | 25270 (2450.41) |
| opt | $Y$ | 1059 (0.94) | 3958 (3.49) | 8139 (22.33) | 25270 (69.19) |
| full | $X \cup Y$ | **768** (9.56) | **2952** (42.83) | **6553** (643.04) | **20526** (2293.2) |
| opt | $X \cup Y$ | **768 (0.77)** | **2952 (3.08)** | **6553 (17.62)** | **20526 (57.52)** |

## 6.2 All Interval Series

Table II compares the different models for finding all the solutions of the all interval series problem. Variable ordering is based on the first-fail principle and value ordering is from the least to the greatest. The models under comparison include the single models: $M_X$ and $M_Y$, the *full* combined model $M_X + C_{\bowtie} + M_Y$, and an *opt*imized combined model IX2 + $C_{\bowtie}$ + IY3 as discussed in Examples 4.12 and 4.15. Puget and Régin [2001] show that all the solutions can be found more efficiently by replacing (IX1.1) and (IX1.2) by (IX1.1') and (IX1.2') respectively where, rather than using a set of separate disequality constraints, we use just two `alldifferent` global constraints

—(IX1.1') `alldifferent`$([x_1, \ldots, x_n])$

—(IX1.2') `alldifferent`$([u_1, \ldots, u_{n-1}])$

The propagator $dsb(\texttt{alldifferent}([v_1, \ldots, v_k]))$ is equivalent to

$$dsb(\wedge_{i=1}^{k-1} \wedge_{j=i+1}^{k} v_i \neq v_j)$$

and has an effective implementation [Régin 1994]. The *pr* model uses IX1.1' + IX1.2' + IX2. The *pr full* model is the combination of *pr* and $M_Y$, IX1.1' + IX1.2' + IX2 + $C_{\bowtie}$ + $M_Y$. The *pr opt* model is the optimized combination of *pr* and $M_Y$, IX1.1' + IX1.2' + IX2 + $C_{\bowtie}$ + IY3 since by Lemma 4.1 the same redundancy reasoning applies.

Compared with the single models $M_X$ and $M_Y$, clearly the *full* and *pr full* models reduce the number of fails significantly (given the same set of search variables as the single models). The *opt* and *pr opt* models maintain the same number of fails as the *full* and *pr full* models respectively for all three set of search variables. Using search variables $Y$, the *opt* model is the fastest for the smaller instance 12 and the *pr full* model is the fastest for larger instances 13, 14 and 15, as the `alldifferent` constraints is computationally too expensive for the smaller instance. Note that the optimized models *opt* and *pr opt* with search variables $Y$ can solve the size 15 instance much faster than *pr*, and no other models can solve this instance within the time limit.

Table II. Results of the All Interval Series Problem

| Model | Search Vars | $n = 12$ fails (sec) | $n = 13$ fails (sec) | $n = 14$ fails (sec) | $n = 15$ fails (sec) |
|---|---|---|---|---|---|
| pr | $X$ | 38778 (24.32) | 156251 (105.26) | 674346 (530.47) | 3045037 (2328.57) |
| $M_X$ | $X$ | 880112 (260.92) | 4914499 (1589.83) | — — | — — |
| full | $X$ | 39241 (222.07) | 158368 (1048.19) | — — | — — |
| opt | $X$ | 39241 (36.34) | 158368 (157.84) | 685301 (770.57) | — — |
| pr full | $X$ | 38461 (236.42) | 155183 (1088.91) | — — | — — |
| pr opt | $X$ | 38461 (42.77) | 155183 (188.94) | 670045 (910.90) | — — |
| $M_Y$ | $Y$ | — — | — — | — — | — — |
| full | $Y$ | 16280 (70.81) | 62949 (303.61) | 266130 (1458.74) | — — |
| opt | $Y$ | 16280 (**6.36**) | 62949 (26.00) | 266130 (108.54) | 1275661 (553.45) |
| pr full | $Y$ | **12296** (62.96) | **43681** (260.90) | **164841** (1127.64) | — — |
| pr opt | $Y$ | **12296** (7.91) | **43681** (**25.78**) | **164841** (**101.42**) | **704097** (**458.12**) |
| full | $X \cup Y$ | 39195 (222.42) | 158282 (1065.77) | — — | — — |
| opt | $X \cup Y$ | 39195 (36.36) | 158282 (158.40) | 684592 (783.01) | — — |
| pr full | $X \cup Y$ | 38447 (230.65) | 155176 (1094.61) | — — | — — |
| pr opt | $X \cup Y$ | 38447 (42.47) | 155176 (198.36) | 669950 (898.66) | — — |

### 6.3 n-Queens Problem

Table III compares the different models for finding all the solutions of the $n$-Queens problem. Variable ordering is based on the first-fail principle and value ordering is from the least to the greatest. The models under comparison include the single models: $M_X$ and $M_Z$, the *full* combined model $M_X + C_\bowtie + M_Z$, and an *opt*imized combined model QX1 + QX2.1 + QX2.2 + $C_\triangle$ + QZ2 as discussed in Examples 4.17 and 4.22. In fact we can show that only the part $\sum_{i=0}^{n-1} z_{ij} \geq 1$ of the (QZ2) constraints in the *opt* model is not propagation redundant, and we refer to this further optimized model as the *part* model.

Although the *full* model with search variables $X$ does reduce the number of fails with respect to the $M_X$ model, but the expensive propagation of the $M_Z$ model overwhelms the benefits. For the other two sets of search variables, the *full* model does not reduce the search space. The *opt* models have the number of fails as the *full* models for all three set of search variables, but we only gain slight improvement in runtime for search variable $X$, we get worse runtime with search variable $Z$ and mixed results with search variable $X \cup Z$. This is because we require more propagation steps to obtain information on the $Z$ variables which are driving the search in both cases. Compared the *part* models against the *opt* models for all three set of search variables, the number of fails are still the same, but the results for runtime are mixed. This is again due to change in propagation steps after the Boolean model is altered. The single model $M_X$ is still the fastest to find all the solutions, followed by the single model $M_Z$. The fact that combining integer model with Boolean model incurred too high runtime overheads, even after removing the redundant propagators, make the combined models impractical. Note that the $n$-queen problem can be solved more efficiently by combining $M_X$ and $M_Y$ using the permutation channels[5] together with a clever value ordering heuristic [Cheng et al.

---

[5]The $M_Y$ model of the $n$-Queens problem has the same set of constraints as $M_X$.

Table III. Results of the $n$-Queens Problem

| Model | Search Variables | $n=11$ fails | (sec) | $n=12$ fails | (sec) | $n=13$ fails | (sec) | $n=14$ fails | (sec) |
|---|---|---|---|---|---|---|---|---|---|
| $M_X$ | $X$ | 21796 | **(2.30)** | 101882 | **(11.27)** | 515238 | **(59.55)** | 2830370 | **(324.95)** |
| full | $X$ | **17601** | (9.43) | **80011** | (45.37) | **392128** | (240.28) | **2101047** | (1350.05) |
| opt | $X$ | **17601** | (8.46) | **80011** | (42.34) | **392128** | (225.30) | **2101047** | (1267.51) |
| part | $X$ | **17601** | (8.46) | **80011** | (42.58) | **392128** | (236.02) | **2101047** | (1273.12) |
| $M_Z$ | $Z$ | 23515 | (3.54) | 111076 | (15.18) | 561362 | (86.12) | 3079792 | (482.95) |
| full | $Z$ | 23515 | (9.63) | 111076 | (50.04) | 561362 | (274.43) | 3079792 | (1591.31) |
| opt | $Z$ | 23515 | (10.40) | 111076 | (53.74) | 561362 | (307.85) | 3079792 | (1715.47) |
| part | $Z$ | 23515 | (10.48) | 111076 | (57.33) | 561362 | (295.14) | 3079792 | (1706.97) |
| full | $X \cup Z$ | 19609 | (9.90) | 90933 | (50.89) | 453886 | (274.82) | 2463621 | (1837.08) |
| opt | $X \cup Z$ | 19609 | (9.93) | 90933 | (48.69) | 453886 | (276.30) | 2463621 | (1521.17) |
| part | $X \cup Z$ | 19609 | (9.56) | 90933 | (49.12) | 453886 | (266.91) | 2463621 | (1685.25) |

Table IV. Results of Social Golfers Problem

| Model | Search Variables | 4-3-4 fails | (sec) | 7-2-13 fails | (sec) | 8-4-9 fails | (sec) | 9-2-17 fails | (sec) |
|---|---|---|---|---|---|---|---|---|---|
| $M_X$ | X | 1509146 | (374.29) | 63860 | (34.51) | **32** | **(0.37)** | **7355** | **(7.89)** |
| full | X | 1509146 | (848.20) | 63860 | (140.83) | **32** | (1.04) | **7355** | (33.04) |
| opt | X | 1509146 | (758.61) | 63860 | (54.81) | **32** | (0.50) | **7355** | (11.55) |
| $M_S$ | S | 2389 | (0.43) | 37158 | (34.91) | — | — | 74098 | (130.72) |
| full | S | **1102** | (0.56) | **27998** | (42.45) | — | — | 51444 | (145.94) |
| opt | S | **1102** | **(0.41)** | **27998** | **(17.74)** | — | — | 51444 | (57.97) |

1999].

### 6.4 Social Golfers Problem

Table IV compares the different models for finding the *first* solution to the social golfers problem. For this problem, we refrain from finding all solutions due to the large number of symmetric solutions. Variable ordering for variables $X$ is based on players by weeks, and variable ordering for variable $S$ is based on weeks by groups. Value ordering is from the least to the greatest for both $X$ and $S$. The models under comparison include the single models: $M_X$ and $M_S$, the *full* combined model $M_X$ + $C_{\{\}}$ + $M_S$, and an *opt*imized combined model GX1 + GX2 + $C_{\{\}}$ as discussed in Examples 4.19 and 4.23. The problem instances use the parameters $g$-$s$-$w$ as described in Example 3.4.

The *full* model with search variables $S$ reduces the number of fails when compare to $M_S$, but not the case when compare to $M_X$ with search variables $X$. The *opt* models speed up the search while keeping the same number of fails as the *full* models for both set of search variables. Using search variables $X$, the search does not benefit from the additional propagation of the combined models *full* and *opt*. In terms of runtime, no one model dominates the others. The *opt* model with search variables $S$ is the fastest for the instances 4-3-4 and 7-2-13, while the single model $M_X$ is the fastest for instances 8-4-9 and 9-2-17.

Table V. Results of Balanced Academic Curriculum Problem

| Model | Search Variables | 8 Periods fails | (sec) | 10 Periods fails | (sec) | 12 Periods fails | (sec) |
|---|---|---|---|---|---|---|---|
| *CPLEX* | n/a | n/a | (1.80) | n/a | (2.27) | n/a | (20.32) |
| *hybrid* | X | **101** | (0.61) | 468 | (2.20) | 58442 | (146.47) |
| *hybrid* | Boolean | 219 | (0.76) | **277** | (1.03) | **315** | (2.09) |
| $M_X$ | X | **101** | **(0.04)** | 468 | (0.25) | 33602 | (11.62) |
| *full* | X | **101** | (0.24) | 470 | (1.80) | 33530 | (192.62) |
| *opt* | X | **101** | (0.08) | 470 | (0.68) | 33530 | (38.54) |
| $M_S$ | S | — | — | — | — | — | — |
| *full* | S | 1577 | (2.83) | 323 | (0.81) | 882 | (4.56) |
| *opt* | S | 1577 | (0.94) | 323 | **(0.24)** | 882 | **(0.95)** |

### 6.5 Balanced Academic Curriculum Problem

Table V compares the model for finding the optimal solution and proving optimality for the problem instances of the balanced academic curriculum problem posted in CSPLib. Variable ordering for $X$ is based on the first-fail principle, and variable ordering for $S$ is based on ascending order of the indices. Value ordering is from the least to the greatest for both set of variables. The models under comparison include the single models: $M_X$ and $M_S$, the *full* combined model $M_X + C_{\{\}} + M_S$, and an *opt*imized combined model B1.1 + B1.2 + B2.1 + B2.2 + BX3 + $C_{\{\}}$ + BS2 + BS3 as discussed in Examples 4.10, 4.19 and 4.24. Hnich *et al.* [2002] report that it is difficult to find the optimal solution and prove optimality with propagation-based constraint solver alone. However, by adding redundant constraints (B2.1) and (B2.2), we were able to solve all the problem instances with $M_X$ alone. The row *CPLEX* gives the runtime for solving the problem instances with ILOG CPLEX 8.0 using an integer linear programming (ILP) model by Hnich *et al.* [2002]. The row *Hybrid* implements the hybrid ILP and CP model described by Hnich *et al.* [2002] together with the redundant constraints (B2.1) and (B2.2) using ILOG Hybrid 1.3.

Clearly, the *full* model with search variables $S$ substantially reduces the search space when compared with the single model $M_S$, but mixed results when compared to the single model $M_X$ with search variables $X$. In solving optimization problems with propagation-based constraint solver, constraints for reducing the bounds of the objective function are added dynamically to reduce the search space. Hence, additional propagation may change the ordering sequence of the solutions, which can increase or decrease the numbers of fails. This explains why the *full* model with search variables $X$ has more fails than the single model $M_X$ for instances of 10 Periods. The *opt* models are faster than the *full* models while keeping the same number of fails for both set of search variables. In terms of runtime, no one model dominates the others. The *opt* model with search variables $S$ is the fastest for solving the instances with 10 and 12 periods, while the $M_X$ model is the fastest for solving the instances with 8 periods. Although the *hybrid* models give the least number of fails, it suffers from the overhead of invoking two solvers.

## 7. RELATED WORK

Smith [2000; 2001] has examined the redundant models for a number of individual problems including the $n$-Queens problem, Langford's problem and the social golfers problems. She empirically demonstrates that some constraints in the redundant models can be removed without increasing the search space. She points out that for these problems the so-called *minimal combined model*, which combine the first model and only the variables of the second model (without the constraints) using channeling constraints, produces the same search behavior as combining the models in full. This is proved in an ad hoc manner by Choi and Lee [2002]. In this paper, we aim for a theoretical framework which can determine propagation redundancy of a particular constraint involved in redundant models *a priori*.

Apt and Monfroy [2001] develop "membership rules" as a way of building propagators for any constraints. Propagation rules are similar to the "membership rules" when restricted to integer variables. However, we develop propagation rules as a method for reasoning about the parts of a propagator's behavior.

Brand [2003] gives a general theorem to determine when a rule is propagation redundant with respect to a set of rules in rule-based constraint programming, and illustrates the applicability using "membership rules." In fact, our definition of a propagation rule satisfies the required properties of Brand's theorem. Hence, we can apply Brand's theorem to determine when a propagation rule is propagation redundant with respect to a set of propagation rules. In this paper, we are interested in propagation redundancy beyond the individual propagation rules, but propagation redundancy of the constraint as a whole. We also generalize the notion of propagation redundancy of a propagation rules through a channel function.

Hnich *et al.* [2004] and Walsh [2001] introduce the notion of *constraint tightness* as a measure to compare the propagation strength of different permutation constraints. Their work focuses on comparing the propagation strength of the different notions of consistency over the disequations, channeling constraints, and `alldifferent` constraints in redundant modeling of *only* permutation problems and injection problems. Our comparison measure is similar to constraint tightness except that constraint tightness is parameterized by a local consistency property. However, in existing constraint solvers, there are propagators which implement *none* of the (established) local consistency properties. An example is the multiplication constraint $x = y \times z$ over integer domain as discussed in Apt [2003, pages 219–220]. In such cases where the local consistency property of a constraint is unknown, our comparison measure would still be applicable. In this paper, we are not only interested in studying the propagation of the permutation constraints, but also the other constraints in redundant models. We also cover a broader class of channeling constraints beyond the permutation channels.

Walsh [2003] proves that "bounds consistency" on set (multiset) variables is equivalent to bounds consistency on the corresponding occurrence representation. This result is related to Theorem 4.26 since the occurrence representation of set variables corresponds to Boolean variables described in Section 3.4. However, existing constraint solvers break Boolean constraints into parts and propagate each part separately. We prove the theorem based on such realistic assumption.

## 8. CONCLUSION

The contributions of this paper are three-fold. First, we define channeling constraints in terms of channel functions which allow us to cover a broad form of redundant modeling. By breaking up a propagator into individual propagation rules, we reason that constraints in one model can be made propagation redundant by constraints in the other model through channels. Second, we introduce the notion of restrictive and unrestrictive channel functions to characterize channeling constraints. Restrictive channel functions can themselves make a constraint in the combined model propagation redundant. Unrestrictive channel functions allow the detection of propagation redundancy of a constraint in one model with respect to a constraint in the other model plus the channels simply in terms of logical consequence. Third, benchmarking results confirm that removals of propagation redundant constraints from combined model can lead to a faster implementation with the same search space. As explained in Section 7, this paper extends related work by covering a broader form of redundant modeling and reasoning about the propagation redundancy of all the constraints in the redundant models.

Our work prompts a number of important future directions for research. It is interesting to investigate if the process of removing propagation redundant constraints can be (semi-)automated. To use Theorem 4.11 we can straightforwardly define the propagation rules for many constraints (parametrically in $D_{init}$) or construct them automatically using the approach of Abdennadher and Rigotti [2002]. The number of propagation rules for most constraints, however, are exponential. A naive implementation is computationally impractical beyond textbook applications. A possible approach is to consider *parameterized propagation rules*, which denotes a set of propagation rules, so that the number of rules is vastly reduced. We can also try to use Theorem 4.20 to prove propagation redundancy without considering propagation rules.

Variable and value ordering heuristics can affect the amount of search space reduction caused by constraint propagation. There are cases shown in Section 6 in which stronger propagation does not reduce the search space. The interesting question is to study a notion of search redundancy which allows even non-propagation redundant constraints to be removed without increasing the search space given specific variable and value ordering heuristics.

Redundant modeling gives rise to the need to decide which variables to label during search. As demonstrated in Section 6, the choice of search variables can greatly affect the size of the search space. Many related work, for example Geelen [1992], Smith [2000; 2001] and Hnich *et al.* [2004], also shown that certain choices of search variables do lead to a smaller search space. Therefore, it is interesting to study and establish criteria in choosing the better set of search variables.